\documentclass[journal, twoside]{IEEEtran}  
\usepackage{import} 

\usepackage[utf8]{luainputenc}

\usepackage[T1]{fontenc}

\usepackage{graphicx}
\usepackage[dvipsnames]{xcolor}

\usepackage[english]{babel}
\addto\captionsenglish{}
\addto\captionsenglish{}
\usepackage{csquotes}

\usepackage{newtxtext}
\usepackage{amsthm}
\usepackage[slantedGreek]{newtxmath}
\usepackage[OMLmathsfit]{isomath}
\DeclareMathAlphabet{\mathbfsf}{\encodingdefault}{\sfdefault}{bx}{n}
\usepackage{bm}
\usepackage{envmath}
\usepackage{mathtools}
\usepackage{commath}

\usepackage[caption=false,font=footnotesize]{subfig}

\usepackage{booktabs}
\usepackage{footmisc}  

\usepackage{url}

\theoremstyle{definition}

\theoremstyle{plain}

\theoremstyle{remark}

\usepackage{lineno}
\modulolinenumbers[5]
\usepackage{todonotes}
\usepackage{umoline}

\usepackage{pgfplots}
\usepackage{pgfplotstable}
\pgfplotsset{compat=newest}
\pgfplotsset{plot coordinates/math parser=false}
\newlength\figureheight
\newlength\figurewidth
\pgfplotsset{every axis plot/.append style={line width=1.5pt},
    legend style={font=\footnotesize, 
        text height=1.0ex,
        draw=black,
        fill=white,
        legend cell align=left}}

\newcounter{subequation}
\newlength\mtabskip\mtabskip=-1.25cm

\def\mtabLong{long}

\newcommand{\mr}{\mathrm}

\newcommand{\veg}[1]{\bm{#1}}     
\newcommand{\mat}[1]{\mathsfbfit{#1}} 
\newcommand{\uv}[1]{\hat{\veg{#1}}} 
\renewcommand{\vec}[1]{\mathsfbfit{#1}} 
\newcommand{\vecop}[1]{\bm{\mathcal{#1}}} 

\newcommand{\jm}{\mathrm{j}}  

\newcommand{\e}{\mathrm{e}}

\newcommand\restr[2]{{
        \left.\kern-\nulldelimiterspace 
        #1 
        \vphantom{|} 
        \right|_{#2} 
}}

\newcommand\rst[3]{{
        \left.\kern-\nulldelimiterspace 
        #1 
        \vphantom{|} 
        \right|_{#2}^{#3} 
}}

\usepackage{acro}

\DeclareAcronym{DG}
{
    short = DG ,
    long = discontinuous Galerkin
}

\DeclareAcronym{ACA}
{
    short = ACA ,
    long = adaptive cross approximation
}

\DeclareAcronym{EFIE}
{
    short =  EFIE ,
    long = electric field integral equation
}

\DeclareAcronym{MFIE}
{
    short =  MFIE ,
    long = magnetic field integral equation
}

\DeclareAcronym{CFIE}
{
    short =  CFIE ,
    long = combined field integral equation
}

\DeclareAcronym{MUIE}
{
    short =  MUIE ,
    long = Müller integral equation
}

\DeclareAcronym{PMCHWT}
{
    short =  PMCHWT ,
    long = Poggio-Miller-Chang-Harrington-Wu-Tsai integral equation
}

\DeclareAcronym{SPD}
{
    short =  SPD ,
    long = {symmetric, positive definite}
}

\DeclareAcronym{SPSD}
{
    short =  SPD ,
    long = {symmetric, positive semi-definite}
}

\DeclareAcronym{PEC}
{
    short =  PEC ,
    long = perfectly electrically conducting
}

\DeclareAcronym{RWG}
{
    short = RWG ,
    long = Rao-Wilton-Glisson
} 

\DeclareAcronym{BC}
{
    short = BC ,
    long = Buffa-Christiansen
}

\DeclareAcronym{SVD}
{
    short = SVD ,
    long = singular value decomposition
}

\DeclareAcronym{CG}
{
    short = CG ,
    long = conjugate gradient
} 

\DeclareAcronym{PCG}
{
    short = PCG ,
    long = preconditioned conjugate gradient
} 

\DeclareAcronym{CGS}
{
    short = CGS ,
    long = conjugate gradient squared
}

\DeclareAcronym{CMP}
{
    short = CMP ,
    long = Calderón multiplicative preconditioner
} 

\DeclareAcronym{RFCMP}
{
    short = RF-CMP ,
    long = refinement-free Calderón multiplicative preconditioner
} 

\DeclareAcronym{HPD}
{
    short = HPD ,
    long = {Hermitian, positive definite}
} 

\DeclareAcronym{RHS}
{
    short = RHS ,
    long = right-hand side
}

\DeclareAcronym{LSE}
{
    short = LSE ,
    long = linear system of equations
} 

\newcolumntype {n}{c}
\newcolumntype {N}{>{\small}c}
\newcolumntype {L}{>{\small}l}
\newcolumntype {F}{>{\footnotesize}c}
\newcolumntype {v}[1]{>{\raggedright \hspace {0pt}} p {#1}}
\newcolumntype {V}[1]{>{\small \raggedright \hspace {0pt}} p {#1}}
\newcolumntype{d}[1]{>{\DC@{.}{.}{#1}}c<{\DC@end}}

\newcolumntype{R}[1]{%
    >{\begin{turn}{90}\begin{minipage}{#1}\small\raggedright\hspace{0pt}}l%
            <{\end{minipage}\end{turn}}%
}

\NewDocumentCommand{\TA}{o}{
    \IfNoValueTF {#1} {%
        \vecop T_{\kern-2pt\mr{A}}
    }
    {
        \vecop T_{\kern-2pt\mr{A},#1}
    }
}

\NewDocumentCommand{\TPhi}{o}{
    \IfNoValueTF {#1} {%
        \vecop T_{\kern-2pt\Phiup}
    }
    {
        \vecop T_{\kern-2pt\Phiup,#1}
    }
}

\NewDocumentCommand{\matTA}{o}{
    \IfNoValueTF {#1} {%
        \mat T_\mr{A}   
        }
    {
        \mat T_{\mr{A},#1}
    }
}

\NewDocumentCommand{\matTPhi}{o}{
    \IfNoValueTF {#1} {%
        \mat T_\Phiup   
        }
    {
        \mat T_{\Phiup,#1}
    }
}

\NewDocumentCommand{\MSL}{o}{
    \IfNoValueTF {#1} {%
        \veg \Psi_\mr{SL}
        }
    {
        \veg \Psi_{\mr{SL},#1}
    }
}

\NewDocumentCommand{\MDL}{o}{
    \IfNoValueTF {#1} {%
        \veg \Psi_\mr{DL}
        }
    {
        \veg \Psi_{\mr{DL},#1}
    }
}

\NewDocumentCommand{\PA}{o}{
    \IfNoValueTF {#1} {%
        \veg \Psi_\mr{A}
        }
    {
        \veg \Psi_{\mr{A},#1}
    }
}

\NewDocumentCommand{\PPhi}{o}{
    \IfNoValueTF {#1} {%
        \veg \Psi_{\Phiup}
        }
    {
        \veg \Psi_{\Phiup,#1}
    }
}

\usepackage{nicefrac}
\usepackage{lipsum} 
\usepackage{shellesc}
\usepackage{multirow}

\usepackage{environ}         
\usepackage{etoolbox}        
\usepackage{graphicx}        
\usepackage{numprint}

\renewcommand{\mat}[1]{\boldsymbol{#1}} 
\renewcommand{\uv}[1]{\hat{\vec{{#1}}}} 
\renewcommand{\vec}[1]{\boldsymbol{#1}}
\newcommand{\pim}[1]{\mathrm{\pi}}
\renewcommand{\jm}{\mathrm{j}}
\newcommand\thefont{\expandafter\string\the\font}

\graphicspath{{./figures/}}  

\usepackage{ifluatex}
\ifluatex

\usepackage{luacode}
\usepackage[strings]{underscore}  

\input{ExternalizingFigures.lua}

\directlua{RecompileAlteredFigures("./figures", "LuaLatex", "MyFigureStyle.sty", 0)}
\fi

\usepackage[author={Matthias Saurer}]{pdfcomment}

\renewcommand{\vec}[1]{\boldsymbol{#1}}
\newcommand{\dyad}[1]{\boldsymbol{\bar{#1}}}

\begin{document}
	
	\title{On the Solution of Linearized Inverse Scattering Problems in Near-Field Microwave Imaging by Operator Inversion and Matched Filtering}
	
	\author{Matthias M. Saurer,~\IEEEmembership{Graduate Student Member,~IEEE},
		Han Na,~\IEEEmembership{Graduate Student Member,~IEEE},\\
		Marius Brinkmann,~\IEEEmembership{Graduate Student Member,~IEEE},
		and Thomas F. Eibert,~\IEEEmembership{Senior Member,~IEEE}
		\thanks{Manuscript received March 1 xxx , 2024; revised October 1 xxx , 2024. This work was funded by the European Union under Grant Agreement No:~101099491.  \it (Corresponding author: Matthias M.~Saurer.) }
		\thanks{Matthias M. Saurer, Han Na, and Thomas F. Eibert are with the Department of Electrical Engineering, School
			of Computation, Information and Technology, Technical University of
			Munich (TUM), 80290 Munich, Germany (e-mail: matthias.saurer@tum.de;
			hft@ei.tum.de). Marius Brinkmann is with Rohde \& Schwarz GmbH \& Co. KG, Munich, Germany (marius.brinkmann@rohde-schwarz.com).}
	}

	\markboth{IEEE TRANSACTIONS ON MICROWAVE THEORY AND TECHNIQUES,~Vol.~XX, No.~X, Januar~20XX}%
	{Saurer \MakeLowercase{\textit{et al.}}: On the Solution of Linearized Inverse Scattering Problems}
	
	\maketitle
	
	\begin{abstract}
		Microwave imaging is commonly based on the solution of linearized inverse scattering problems by matched-filtering algorithms, i.e., by applying the adjoint of the forward scattering operator to the observation data. A more rigorous approach is the explicit inversion of the forward scattering operator, which is performed in this work for quasi-monostatic imaging scenarios based on a planar plane-wave representation according to the Weyl-identity and hierarchical acceleration algorithms. The inversion is achieved by a regularized iterative linear system of equations solver, where irregular observations as well as full probe correction are supported. In the spatial image generation low-pass filtering can be considered in order to reduce imaging artifacts. A corresponding spectral back-projection algorithm and a spatial back-projection algorithm together with improved focusing operators are also introduced and the resulting image generation algorithms are analyzed and compared for a variety of examples, comprising both simulated and measured observation data. It is found that the inverse source solution generally performs better in term of robustness, focusing capabilities, and image accuracy compared to the adjoint imaging algorithms either operating in the spatial or spectral domain. This is especially demonstrated in the context of irregular sampling grids with non-ideal or truncated observation data  and by evaluating all reconstruction results based on a rigorous quantitative analysis.
		
	\end{abstract}
	
	
	\section{Introduction}
	\IEEEPARstart{T}{he} great benefits of microwave radiation up to the mm-wave bands are its non-ionizing character but still very good resolution capabilities as well as its potential of penetrating through clothing barriers and dielectric walls~\cite{Sheen.Sep.2001,Ahmed.Sep.2012}. Due to these properties, microwave-based imaging techniques are used in various areas such as medical imaging~\cite{Tajik.,Gilmore.Jun.2009}, satellite remote sensing~\cite{LeVine.Dec.1999,Krieger.May2014}, or non-destructive testing~\cite{Lopez.Jun.2022,Brinkmann.Oct.2022}. In particular the concept of synthetic aperture radar (SAR) has been proven as a versatile and reliable method in this regard~\cite{Saurer.Oct.2022,Na.Oct.2023}. In SAR, the movement of a carrier platform, e.g., an airplane carrying the transmit and receive antennas is utilized in order to synthesize larger virtual apertures, which consequently lead to an increase of resolution in cross-range direction~\cite{Moreira.Mar.2013}. It is clear that this technique is conceptually interesting for various applications such as drone measurements~\cite{Punzet.May2022}, automotive radar~\cite{Farhadi.Apr.2022}, or freehand smartphone imaging~\cite{AlvarezNarciandi.Mar.2021}. However, for a successful adaption of SAR imaging to these applications, different challenges have to be overcome. This is mainly due to the fact that the data acquisition for these examples is subject to  irregular sampling. The realization of flexible and computationally inexpensive reconstruction algorithms, which can handle the irregularities, is, thus, a task of increasing significance~\cite{Smith.Jan.2022}. 
	
	Mathematically, radar and imaging algorithms rely commonly on the solution of a linearized inverse scattering problem, where, according to the first-order Born approximation, independent scattering center distributions are assumed, which only interact with the incident field~\cite{Haynes.Jan.2024}.
	This linearization of the inverse source problem allows to express the scattered electric field as a convolution of the scattering center distribution of the target with the dyadic Green's function of the underlying homogeneous space~\cite{Schnattinger.May2012}. Consequently, employing a plane-wave expansion leads to a diagonalization of the linearized integral operator and, hence, in theory a direct inversion of the inverse source problem is possible. However, even within the first-order Born approximation the forward operator exhibits a null-space~\cite{Devaney.2012} leading to uniqueness problems, and the efficient processing of irregularly distributed observation locations and arbitrary probe orientations still poses a great challenge. 
	
	 The corresponding imaging algorithms can be divided into direct reconstruction methods by adjoint imaging, sometimes also referred to as matched filtering~\cite{Dyab.Oct.2013}, and operator inversion algorithms. Regarding direct image reconstruction, the standard back-projection algorithm (BPA) \cite{Ahmed.Apr.2014,Batra.Apr.2021}, also known as delay and sum method~\cite{Alkhodary.May2016}, is widely used mainly due to its robustness and ease of implementation. Computationally much more efficient than the BPA are Fourier transform based $\omega$-$k$-methods \cite{Smith.Jan.2022,Saurer.Jul.2022}, which perform the back-projection on a plane-wave basis in the spatial frequency domain and which can consider spectral filtering functions in a straightforward way. For efficient evaluation together with regular observation grids, they rely typically on fast Fourier transforms (FFTs). An efficient treatment of irregular observation grids is, e.g., possible with non-uniform FFTs~\cite{Wang.Jan.2020} or utilizing the multi-level fast spectral domain algorithm (MLFSDA) in~\cite{Saurer.Jul.2022}, which exhibits excellent computational efficiency due to the use of hierarchical multi-level schemes as employed in the multi-level fast multipole method (MLFMM)~\cite{Chew.2001}. Common operator inversion methods start with an initial guess of the scattering distribution directly in the spatial domain, which is then further improved in terms of noise suppression or contrast enhancement by iteratively minimizing a least mean squares cost functional~\cite{Kelly.Jul.2011,Roberts.Feb.2010}. Alternatively, one may work with a spectral propagating plane-wave decomposition of the forward operator, which allows to solve rather well-conditioned and very compact single-frequency inverse source problems~\cite{Schnattinger.Aug.2014,Neitz.May2019}. The solution process can here be effectively accelerated by hierarchical concepts as found in the MLFMM and the spatial images are created by coherent superposition of the single-frequency images, which may, e.g., be obtained by hierarchical disaggregation~\cite{Schnattinger.May2012}.
	While it is clear that operator inversion methods provide an  estimate of the solution quality by evaluating the norm of the residual error vector, the iteration process itself is of course more time- and memory consuming than the adjoint imaging methods. 
	
	Dependent on the utilized field representation and the imaging configuration, it can occur that the adjoint operator is identical to the inverse operator. Formally, this can be achieved by diagonalization and normalization of the scattering operator, e.g., in the form of a plane-wave or spherical mode expansion of the observations. However, in practical configurations this is typically not feasible due to truncation effects, sampling implications, and probing antenna influences. 
	Another way of improving the adjoint imaging methods is to work with modified imaging operators, which are, e.g., calibrated via the obtained point spread functions \cite{Osipov.Sep.2013,Broquetas.May1998,Watanabe.Mar.2022}. 
	
	The goal of this article is to study and demonstrate the imaging properties of operator inversion based algorithms and contrast them to the corresponding properties of adjoint imaging methods, where the focus is on practically relevant planar and quasi-planar imaging configurations. The utilized operator inversion method is working according to the concepts of the MLFSDA based $\omega$-$k$-algorithm~\cite{Saurer.Jul.2022} for the operator evaluations within the iterative solver and has, thus, the ability to handle irregular observation sample distributions and arbitrary probing antennas.
	For the adjoint imaging, the MLFSDA based $\omega$-$k$-algorithm is employed as it is, but a standard spatial BPA is used as well, where in particular different focusing operators are derived from the plane-wave based adjoint operator representation. 
	The superiority of the operator inversion method over the direct reconstruction approaches is in particular demonstrated for irregular observation locations, where simulated and measured observation data is used. Moreover, it is shown that spectral low-pass filtering functions can effectively suppress imaging artifacts in all of the considered algorithms.
	
	 The rest of this article is organized as follows. The formulation and derivation of the various imaging algorithms are given in Section~\ref{ISCP}. The obtained imaging results by utilizing simulated as well as measured observation data are presented and thoroughly discussed in Section~\ref{sim_setup} and Section~\ref{meas}, respectively. Based on this, some conclusions are drawn in Section~\ref{concl}.
	
	\section{Formulation of the imaging algorithms}\label{ISCP}
	\subsection{Problem Statement and Planar Plane-Wave Representation}
\begin{figure}
		\centering
		\includegraphics[scale=0.70]{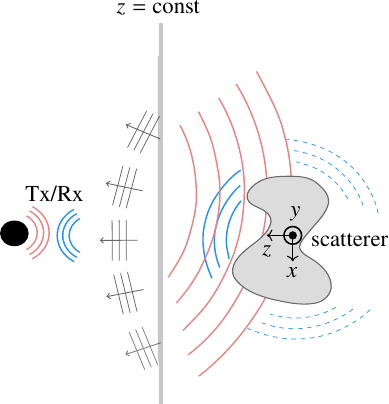}\hfill
		\caption{Visualization of the imaging scenario. The scattering behavior of an unknown object is equivalently described by utilizing a planar plane-wave expansion in the Huygens-plane $z=\mathrm{const}$.}
		\label{vis_Huygens_plane}
	\end{figure}

\begin{figure}[t]
		\centering
		\includegraphics[scale=0.8,keepaspectratio]{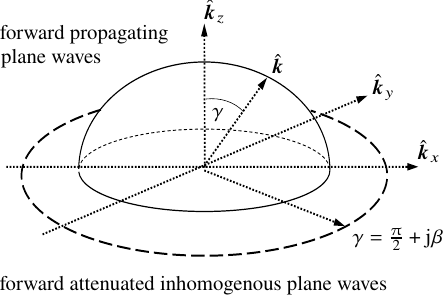}
		\caption{Integration contour $\Gamma$ in $k$-space for the angular spectrum evaluation of the Weyl-identity~(\ref{weyl_ang}). The integration area is covered by integrating $\gamma$ from 0 to $\uppi/2$ and from $\uppi/2$ to $\uppi/2+\mathrm{j}\infty$ for a full $2\uppi$-integration of the rotation angle around the $\hat{\vec{k}}_z$-axis. The figure is slightly adapted from~\cite{Eibert.Jan.2024}.} 
		\label{gamma_weyl}
	\end{figure}

	Consider a quasi-monostatic scattering scenario, where the transmitting antenna (Tx), generating the incident fields, and the receiving antenna (Rx), recording the scattered fields, are either collocated or very close to each other at observation positions $\vec{r}_m$ with $m = 1,2,\dots,M$. As depicted in Fig.~\ref{vis_Huygens_plane}, it is assumed that the scattering centers and the Tx/Rx locations are well separated, thus, allowing to introduce a planar Huygens' surface in between, which is well suited to define the local support of an equivalent plane-wave representation of the incident and scattered fields. 
	Assuming the validity of the first-order Born approximation, the transfer operator between the two antennas via the scatterer can be written as~\cite{Schnattinger.Aug.2014}
	\begin{equation}
		T(\vec{r}_m)=
		\iiint\limits_{V_{\mathrm{S}}}
		\vec{E}_{\mathrm{R}}\left({\vec{r}},\vec{r}_{m}\right)
		\cdot
		\dyad{s}_{\mathrm{B}}\left({\vec{r}}\right)
		\cdot
		\vec{E}_{\mathrm{T}}\left({\vec{r},\vec{r}_{m}}\right)
		\,
		\mathrm{d}^3 \vec{r}\,,
		\label{scat_spat}
	\end{equation}
	where $\dyad{s}_{\mathrm{B}}\left({\vec{r}}\right)$ is the scattering dyad defined in the volume $V_{\mathrm{S}}$ at the position vector $\vec r = (x,y,z)$ with $r=\abs{\vec{r}}$, $\vec{E}_{\mathrm{T}}\left({\vec{r},\vec{r}_{m}}\right)$ is the radiated electric field vector of the Tx antenna, and $\vec{E}_{\mathrm{R}}\left({\vec{r}},\vec{r}_{m}\right)$ is the radiated  electric field vector of the Rx antenna, which is reciprocal to the scattered field received by the Rx antenna. 
	In order to arrive at a spectral representation of the transfer operator, we utilize the Weyl-identity in its original angular plane-wave spectrum form~\cite{H.Weyl.1919,Booker.1950}
	\begin{equation}\label{weyl_ang}
		\frac{\e^{-\mathrm{j}kr}}{r}=\frac{k}{2\uppi\jm}\iint\limits_{\Gamma}^{}\e^{-\mathrm{j}k_z \abs{z}}\e^{-\mathrm{j}\left(k_x x+k_y y\right)}\,\mathrm{d}^{2} \hat{\vec{k}}\,
	\end{equation}
	with an integration contour $\Gamma$ as explained and illustrated in Fig.~\ref{gamma_weyl}, where the unit vector $\uv{k}$ is oriented into the direction of the wave vector $\vec{k}=(k_x,k_y,k_z)$ with $k=|\vec{k}|$.
	Assuming moreover a time dependence $\e^{\,\jm\omega t}$ with angular frequency $\omega$, the radiated antenna fields incident on the scatterer can be expressed as~\cite{Saurer.Jul.2022,Kong.1990,Eibert.2015}
	\begin{equation}
		\vec{E}_{\mathrm{T/R}}\left({\vec{r},\vec{r}_{m}}\right)=\iint\limits_{\Gamma^{\rm i}}^{}\tilde{\vec{W}}_{\mathrm{T/R}}\left(\vec{k}^{\rm i}\right)\e^{-\jm \vec{k}^{\rm i}\cdot\left(\vec{r}-\vec{r}_m\right)}\,\mathrm{d}^{2} \hat{\vec{k}}^{\rm i}\,\label{pw_inc}
	\end{equation}
	in terms of the angular plane-wave expansions $\tilde{\vec{W}}_{\mathrm{T}}\left(\vec{k}^{\rm i}\right)$ and $\tilde{\vec{W}}_{\mathrm{R}}\left(\vec{k}^{\rm i}\right)$ of the Tx and Rx antennas, respectively, which may be obtained from
	\begin{equation}
		\tilde{\vec{W}}_{\mathrm{T/R}}\left(\vec{k}^{\rm i}\right)=\frac{-Zk^2 }{8\uppi^2}\left(\mathbf{\bar{I}}-\hat{\vec{k}}^{\rm i}\hat{\vec{k}}^{\rm i}\right)\cdot\iiint\limits_{V_{\mathrm{T/R}}}\vec{J}_{\mathrm{T/R}}\left(\vec r'\right)\e^{\,\jm \vec{k}^{\rm i}\cdot\vec{r}'}\,\mathrm{d}^3 \vec{r}',\label{wtr}
	\end{equation}
	if $\vec{J}_{\mathrm{T/R}}(\vec r')$ are corresponding  spatial equivalent electric current densities producing the antenna radiation in free space. $Z=\sqrt{\mu/\varepsilon}$ is the wave impedance and $|\vec{k}^{\rm i}|= \omega\sqrt{\varepsilon\mu}$ the necessary dispersion relation, where $\varepsilon$ and $\mu$ are the permittivity and the permeability of the homogeneous background space, respectively. 
	Assuming propagating plane waves towards the scatterer only, i.e., $\gamma$ on the integration contour shown in Fig.~\ref{gamma_weyl} ranging from 0 to $\uppi/2$, $\tilde{\vec{W}}_{\mathrm{T}}\left(\vec{k}^{\rm i}\right)$ and $\tilde{\vec{W}}_{\mathrm{R}}\left(\vec{k}^{\rm i}\right)$ correspond to the far-field patterns of the antennas. These probe patterns can be obtained by measurements or by full-wave simulations of the antennas~\cite{Eibert.2015}. 
	Typically both far-field patterns are expressed in spherical vector components according to
	\begin{equation}
		\tilde{\vec{W}}_{\mathrm{T/R}}\left(\vec{k}^{\rm i}\right)= \tilde{W}_{\mathrm{T/R}}^{\vartheta}\left(\vec{k}^{\rm i}\right)\uv{\vartheta}+
		\tilde{W}_{\mathrm{T/R}}^{\varphi}\left(\vec{k}^{\rm i}\right)\uv{\varphi}\,,
	\end{equation}
	where $\uv{\vartheta}$ and $\uv{\varphi}$ are the corresponding spherical unit vectors.
	Plugging (\ref{pw_inc}) into (\ref{scat_spat}) and assuming the radiating reflectors model or correspondingly that the scattering process is purely  monostatic in the $k$-space~\cite{Schnattinger.Aug.2014,Claerbout.1985}, reduces the 4D bistatic integral into a monostatic 2D integral and results in a spectral representation of the near-field transfer operator for polarimetric imaging in the form 
	\begin{equation}\label{gen_nf0}
		T\left(\vec{r}_m\right) =\iint\limits_{\Gamma}^{}\vec{\tilde{W}}_{\mathrm{R}}\left(-\vec{k}\right)\cdot\tilde{\dyad{s}}_{\mathrm{B}}\left(2\vec{k}\right)\cdot\vec{\tilde{W}}_{\mathrm{T}}\left(-\vec{k}\right)\e^{\,-\jm 2 \vec{k}\cdot\vec{r}_m}\,\mathrm{d}^{2} \hat{\vec{k}}\,,
	\end{equation}
	where $\vec k = -\vec k^{\rm i}$ has been introduced and where $\tilde{\dyad{s}}_{\mathrm{B}}\left(2\vec{k}\right)$ is the spectral scattering distribution defined as
	\begin{equation}
		\tilde{\dyad{s}}_{\mathrm{B}}\left(2\vec{k}\right) =\iiint\limits_{V_{\mathrm{S}}}
		\dyad{s}_{\mathrm{B}}\left(\vec{r}\right)
		\e^{\,\jm 2 \vec{k}\cdot\vec{r}}\,\mathrm{d}^3 \vec{r}\,.
		\label{sks}
	\end{equation}
	Mapping of the integration contour $\Gamma$ of the Weyl-identity in~(\ref{weyl_ang}) into the $(k_x,k_y)$-plane~\cite{Booker.1950}, results into its probably even more common planar plane-wave spectrum form~\cite{Kong.1990}
	\begin{equation}\label{weyl_plane}
		\frac{\e^{-\mathrm{j}kr}}{r}=\iint\limits_{-\infty}^{+\infty}\frac{\e^{-\mathrm{j}k_z \abs{z}}}{2\uppi\mathrm{j}k_z}\e^{-\mathrm{j}\left(k_x x+k_y y\right)}\, \mathrm{d}{k}_x\,\mathrm{d}{k}_y\,.
	\end{equation}
	Applying the same mapping to the near-field transfer or forward operator in~(\ref{gen_nf0}) gives finally the planar plane-wave representation of the near-field transfer operator
	\begin{equation}\label{gen_nf1}
		T\left(\vec{r}_m\right) =\iint\limits_{-\infty}^{+\infty}\vec{\tilde{W}}_{\mathrm{R}}\left(-\vec{k}\right)\cdot\tilde{\dyad{s}}_{\mathrm{B}}\left(2\vec{k}\right)\cdot\vec{\tilde{W}}_{\mathrm{T}}\left(-\vec{k}\right)\frac{\e^{\,-\jm 2 \vec{k}\cdot\vec{r}_m}}{k k_z}\,\mathrm{d}{k}_x\,\mathrm{d}{k}_y,
	\end{equation}
	which shall be the basis of our subsequent considerations. In literature, this equation is sometimes found with a $k_z^2$ in the denominator instead of the $k_z$. This may happen, if the derivation is carried out with the incident field~(\ref{pw_inc}) directly given in a planar plane-wave representation according to the Weyl-identity in~(\ref{weyl_plane}), instead of (\ref{weyl_ang}). According to the concept of the angular spectrum of plane waves~\cite{H.Weyl.1919,Booker.1950}, this is not correct. However, in terms of imaging, such a representation leads often even to better images, since the compensation of the $k_z^2$ in the denominator during the image generation is equivalent to a low-pass filtering of the image.
	In order to handle different Tx/Rx probing antenna combinations in a numerical implementation, the transfer or forward operator~(\ref{gen_nf1}) is rewritten in a column vector format according to
	\begin{equation}\label{gen_nf}
		T_p\left(\vec{r}_m\right) =\iint\limits_{-\infty}^{+\infty}\mat{\tilde{W}}_p\left(-\vec{k}\right)\cdot\tilde{\vec{S}}_{\mathrm{B}}\left(2\vec{k}\right)\frac{\e^{\,-\jm 2 \vec{k}\cdot\vec{r}_m}}{k k_z}\,\mathrm{d}{k}_x\,\mathrm{d}{k}_y\,,
	\end{equation}
	where the column vector 
	$\vec{\tilde{W}}_p\left(-\vec{k}\right)$ contains the spectral probe weighting coefficients for the Tx/Rx probe combination $p$ with \mbox{$p=1,2,\dots,P$} and the column vector $\tilde{\vec{S}}_{\mathrm{B}}(2\vec{k})$ contains the spherical scattering components of the dyad in~(\ref{sks}). 
	The 4D column vector expressions are given as
	\begin{equation*}
		\vec{\tilde{W}}_{p}\left(\vec{k}\right) = \left[\begin{array}{c}
			\tilde{W}_{\mathrm{R}_p}^{\vartheta}\left(\vec{k}\right)\tilde{W}_{\mathrm{T}_p}^{\vartheta}\left(\vec{k}
			\right)\vspace{0.1 cm}\\
			\tilde{W}_{\mathrm{R}_p}^{\varphi}\left(\vec{k}\right)\tilde{W}_{\mathrm{T}_p}^{\varphi}\left(\vec{k}\right)\vspace{0.1cm}\\
			\tilde{W}_{\mathrm{R}_p}^{\varphi}\left(\vec{k}\right)\tilde{W}_{\mathrm{T}_p}^{\vartheta}\left(\vec{k}\right)\vspace{0.1 cm}\\
			\tilde{W}_{\mathrm{R}_p}^{\vartheta}\left(\vec{k}\right)\tilde{W}_{\mathrm{T}_p}^{\varphi}\left(\vec{k}\right)
		\end{array}\right],\,\, \tilde{\vec{S}}_{\mathrm{B}}\left(2\vec{k}\right) = \left[\begin{array}{c}
			\tilde{s}_{\vartheta\vartheta}\left(2\vec{k}\right)\\
			\tilde{s}_{\varphi\varphi}\left(2\vec{k}\right)\\
			\tilde{s}_{\varphi\vartheta}\left(2\vec{k}\right)\\
			\tilde{s}_{\vartheta\varphi}\left(2\vec{k}\right)
		\end{array}\right]\,.
	\end{equation*}
	Utilizing the orthogonality of the plane waves in the observation plane $z=z_m$, the spectral scattering distribution is obtained from (\ref{gen_nf}) according to
	\begin{equation}\label{gen_inf0}
		\mat{\tilde{W}}_p\left(-\vec{k}\right)\cdot\frac{\tilde{\vec{S}}_{\mathrm{B}}\left(2\vec{k}\right)}{k k_z}=\frac{1}{\uppi^2}\iint\limits_{-\infty}^{+\infty}T_p\left(\vec{r}_m\right)\\\e^{\,\mathrm{j}2\vec{k}\cdot\vec{r}_m}\,\mathrm{d}x_m\,\mathrm{d}y_m
	\end{equation}
	and the spatial scattering distribution is finally derived in the planar plane-wave representation 
	\begin{align}
		\vec{S}_{\mathrm{B}}(\vec{r}) & = \frac{1}{\uppi^2}
		\iint\limits_{-\infty}^{+\infty}H_n(2\vec{k})\frac{\tilde{\vec{S}}_{\mathrm{B}}\left(2\vec{k}\right)}{kk_z}\e^{\,-\mathrm{j}2\vec{k}\cdot\vec{r}}\,\mathrm{d}{k}_x\,\mathrm{d}{k}_y\label{img_gen}\\
		&=\frac{1}{\uppi^4}\iint\limits_{-\infty}^{+\infty}H_n(2\vec{k})\left[\vec{\tilde{W}}\left(-\vec{k}\right)\right]^{\dagger}\cdot
		\iint\limits_{-\infty}^{+\infty}\vec{T}\left(\vec{r}_m\right)\notag\\&\phantom{~~~~~~~~~~~~~~~~} \e^{\,\mathrm{j}2\vec{k}\cdot\vec{r}_m}\,\mathrm{d}x_m\,\mathrm{d}y_m \;\;\e^{\,-\mathrm{j}2\vec{k}\cdot\vec{r}}\,\mathrm{d}{k}_x\,\mathrm{d}{k}_y\,,\label{gen_inf}
	\end{align}
	where $\vec{S}_{\mathrm{B}}(\vec r)$ is the column vector representation of the scattering dyad $\dyad{s}_{\mathrm{B}}(\vec r)$, $\vec{T}\left(\vec{r}_m\right)$ contains the scattering measurements obtained for different Tx/Rx probe antenna combinations in one vector, and $H_{n}(2\vec{k})$ represents a low-pass filtering function, which can be used to mitigate artifacts due to the truncation of the plane-wave spectrum or the observation area~\cite{Devaney.2012}.
	$\vec{\tilde{W}}\left(-\vec{k}\right)$ denotes a probe correction matrix, which is obtained by stacking the vectors $\vec{\tilde{W}}_p\left(-\vec{k}\right)$ for $p=1,...,P$ row-wise in a matrix. Since this matrix can be severely ill-conditioned dependent on $\vec{k}$, its pseudo-inverse denoted by the symbol $\dagger$ is used in~(\ref{gen_inf}). This equation is the basis for the MLFSDA based $\omega$-$k$-algorithm as introduced in~\cite{Saurer.Jul.2022}, where the final image is obtained by coherently summing up all the single-frequency images according to~(\ref{gen_inf}) for all considered frequencies. Due to its explicit inversion in the spectral domain,~(\ref{gen_inf}) can provide more accurate results than a standard adjoint operator evaluation. However, it is strictly correct only for measurements available in an infinite observation plane and for sets of Tx/Rx probe combinations with the same plane-wave spectrum in every measurement location~$\vec{r}_m$. In practical implementations, the spatial integral over the observations in $\vec{T}(\vec{r}_m)$ is often just performed as a discrete sum over the available observations where even 3D irregular observation locations may be considered~\cite{Gumbmann.Apr.2017}. Complications in the evaluation of~(\ref{gen_inf}) might arise from the computation of the pseudo-inverse of $\tilde{\vec{W}}(-\vec{k})$, which can be strongly rank deficient.  Since the MLFSDA based $\omega$-$k$-algorithm assumes uniform weighting factors to compute the plane-wave spectra of the measurements, it is for arbitrary sampling grids and for truncated observation areas in general more error prone compared to the proposed operator inversion approach. Additionally, instead of computing the pseudo-inverse of the probe matrix according to (\ref{gen_inf}), the inverse source based approach employs a multiplication of the probe coefficients within its operator evaluations and, hence, is numerically more stable. For regularly distributed sampling points, (\ref{gen_inf}) can be solved very efficiently also using 2D-FFT based $\omega$-$k$ methods~\cite{Smith.Jan.2022}.

	The low-pass filtering function $H_{n}(2\vec{k})={\left(2k_z\right)}^n$ with $n\in\mathbb{N}_0$ as introduced in~(\ref{img_gen}) and~(\ref{gen_inf}) can help to improve the image quality in all $\omega$-$k$ based methods. As will be shown later, the chosen form of filtering function allows for a closed-form representation of the focusing operator when utilizing the spatial back-projection algorithm. Since such a filtering operation reduces the bandwidth in the spatial frequency domain and, thus, also the achievable resolution in the reconstructed images, it should be used with care. In literature, such techniques are known as filtered back-projection methods~\cite{Devaney.2012}.
	
	\subsection{Iterative Operator Inversion}
	The operator inversion approach is starting from the near-field transfer operator in its spatial frequency form as given in~(\ref{gen_nf}). In order to avoid numerical instabilities at the boundary of the visible region, it is convenient to define the auxiliary spectral scattering vector $\tilde{\vec{S}}_{\mathrm{B,inv}}(2\vec{k})$ = $\tilde{\vec{S}}_{\mathrm{B}}(2\vec{k})/(kk_z)$, which is directly the  quantity required to generate the image according to (\ref{img_gen}). 
	More specifically,
	the plane-wave expansion coefficients $\tilde{\vec{S}}_{\mathrm{B,inv}}(2\vec{k}_n)$ of the scattering object according to
	\begin{equation}\label{disk_s}
	\tilde{\vec{S}}_{\mathrm{B,inv}}(2\vec{k})=\sum_{n=1}^{N}\tilde{\vec{S}}_{\mathrm{B,inv}}(2\vec{k}_n)\delta\left(2\vec{k}-2\vec{k}_n\right)\,
	\end{equation}
    are considered as the unknown sources of the corresponding inverse source problem, where $\delta(2\vec{k})$ denotes the Dirac-delta distribution and $2\vec{k}_n$ with $n=1,\dots,N$ corresponds to the $n$th plane-wave sample in the spatial frequency domain according to a regular grid in $({k}_x, k_y)$. The resulting unknown spherical coefficients are collected in the unknown source vector $\mat{x}\in\mathbb{C}^{N}$.
	The $M$ observation samples according to $T_p(\vec{r}_m)$ are stacked into the measurement vector $\mat{b}\in\mathbb{C}^{M}$. By defining the matrix $\mat{A}\in\mathbb{C}^{M\times N}$, which is the discrete numerical representation of the near-field transfer operator according to (\ref{gen_nf}), a linear system of equations is obtained in the form $\mat{A}\vec{x}=\vec{b}$. Since this system of equations is typically at least mildly ill-posed and may have less or more unknowns than equations, it is solved in the form of the normal error system of equations 
	\begin{equation}\label{ne_eq}
		\mat{A}\mat{A}^{\mathrm{ad}}\,\mat{u} = \mat{b}\,,	
	\end{equation}
	where $\mat{x}=\mat{A}^{\mathrm{ad}}\,\mat{u}$ has been introduced with $^\mathrm{ad}$ indicating the adjoint operator.
	This systems of equations is solved by the iterative generalized minimal residual (GMRES) solver~\cite{Saad.1986}, which does not require an explicit matrix formation and is, thus, especially beneficial for large-scale problems. The GMRES solver utilizes a Krylov sub-space of increasing size during the iterative solution process, which resembles a power series evaluation with respect to the matrix operator $\mat{A}\mat{A}^{\mathrm{ad}}$. Utilizing the iterative GMRES method, therefore, effectively suppresses non-radiating sources and overcomes the problem of non-uniqueness by generating a minimum norm solution vector $\vec{u}$, which is inherently regularized. Moreover, working with the normal error system of equations is beneficial in the sense that it allows for a direct monitoring of the observation error~\cite{Kornprobst.2019,Kornprobst.Aug.2021}. Hence, the iterative solution process can be directly stopped, once the normalized observation error is below a threshold $\varepsilon$ or the ratio of the normalized observation errors of two consecutive iterations is above a relative threshold  $\varepsilon_{\mathrm{rel}}$~\cite{Ostrzyharczik.Feb.2023}. During the iterative solution process it is mandatory that the forward and adjoint operators are efficiently evaluated and this is achieved by following the concepts of the MLFSDA as presented in~\cite{Saurer.Jul.2022}, which are of course again related to the concepts of the MLFMM~\cite{Chew.2001}. For the evaluation of the forward operator according to~(\ref{gen_nf}), first the primary sources on the finest level are interpolated by an exact FFT-based global interpolation scheme~\cite{Sarvas.2003} in order to compute plane-wave spectra with an oversampling factor of two. The sub-spectra on the finest source level are then 
	aggregated to an overall source spectrum of plane waves by an hierarchical aggregation scheme utilizing local Lagrange interpolations.
	
	Next, the translation operator $T(\vec{r}_m,2\vec{k})=\exp(-\jm 2\vec{k}\cdot\vec{r}_m)$ is multiplied to the $k$-space samples of the scattering distribution, where the translation occurs first into a reference location within the observation plane and the translations into the individual measurement locations $\vec{r}_m$ are finalized during the subsequent hierarchical disaggregation and down-sampling scheme. Finally the influence of the probing antennas is considered by taking the inner product of the plane-wave samples representing the scattering object with the probe weighting coefficients contained in $\tilde{\vec{W}}_p(-\vec{k})$ and the $k$-space integral is evaluated by summing up all plane-wave samples multiplied with the corresponding quadrature weights.
	Since the required sampling rate for the representation of the translation operator is proportional to the maximum translation distance, it can be of benefit to utilize spatial filtering strategies as discussed in \cite{Eibert.Jan.2024} in order to reduce the required sample density according to the sizes of the considered source and observation areas. For the evaluation of the adjoint operator, the described process is just performed in reverse order and by considering the conjugate complex of any involved complex coefficients. 
	
	In the iterative solution of (\ref{ne_eq}),  the joint handling of observation samples obtained with different probing antenna combinations or different probe orientations does not cause any problem, since all the available observation samples are just collected in the vector $\vec{b}$. In contrast, the MLFSDA based $\omega$-$k$-method according to~(\ref{gen_inf}) can only transform observation samples for sets of identical probing antenna combinations and  orientations, where probe correction is performed in the spectral domain after combining different transformation results as shown in~\cite{Saurer.Jul.2022}. This, however, poses a limitation in terms of flexibility and robustness. The described iterative solution of the inverse source problem requires obviously a larger total amount of computational operations than an adjoint imaging method, but it also provides an error estimate that indicates how well the found solution fits to the given observation data. Due to the factorized form of the forward and adjoint operators, the inverse source based imaging algorithm achieves the same numerical complexity of $\mathcal{O}(N\log N)$ per iteration as the MLFSDA based $\omega$-$k$-algorithm for the computation of the plane-wave spectra. After obtaining the spectral scattering vector $\vec{\tilde{S}}_{\mathrm{B,inv}}(2\vec k)$ the spatial images are generated by evaluating~(\ref{img_gen}),
	which can be implemented very efficiently using inverse FFTs or a hierarchical disaggregation scheme as discussed in~\cite{Schnattinger.Nov.2012}. 

	\subsection{Improved Focusing Operators for the Monostatic BPA}
	\label{sbpa}
	Starting from the spatial transfer operator as given in~(\ref{scat_spat}), a spatial adjoint monostatic back-projection equation for observation samples in a plane $z=z_m$ may formally be written according to
	\begin{equation}
		\dyad{s}_{\mathrm{B}}\left({\vec{r}}\right)=
		\iint\limits_{-\infty}^{+\infty}
		\vec{E}^{\mathrm{ad}}_{\mathrm{R}}\left({\vec{r}},\vec{r}_{m}\right)
		T(\vec{r}_m)\,
		\vec{E}^{\mathrm{ad}}_{\mathrm{T}}\left({\vec{r},\vec{r}_{m}}\right)
		\,
		\mathrm{d} x_m\,\mathrm{d} y_m\,,
		\label{bpab}
	\end{equation}
	where the vectors in the integral are multiplied in the form of a dyadic product.
	{In a scalar wave formulation with the assumption of a scalar scattering distribution in the far field of the probing antennas with isotropic patterns and ignoring some of the constants, this equation reduces to
 \begin{equation}
    	{s}_{\mathrm{B}}\left({\vec{r}}\right)=
    	\iint\limits_{-\infty}^{+\infty}
    	T(\vec{r}_m)\,\e^{\,\mathrm{j}2k\abs{\vec{r}_m-\vec{r}}}\,
    	\mathrm{d} x_m\,\mathrm{d} y_m\,,
    	\label{naive_bpa}
    \end{equation}	
	which is recognized as a standard back-projection equation~\cite{Ozdemir.2021}. Recognizing that the direct imaging equation~(\ref{gen_inf}) is based on an operator inversion in the spatial frequency domain, it can be expected that it produces improved imaging results as compared to~(\ref{naive_bpa}) and that this can be the starting point for deriving an improved BPA for planar (or quasi-planar) observation configurations. By changing the integration order in~(\ref{gen_inf}), a spectral representation of the spatial focusing operator is obtained, where in particular also the spectral filtering function $H_{n}(2\vec{k})={\left(2k_z\right)}^n$ can be considered. For the scalar case according to~(\ref{naive_bpa}), this results in
	\begin{equation}\label{foc_def}
		s_{\mathrm{B}}(\vec{r})=\iint\limits_{-\infty}^{+\infty}T\left(\vec{r}_m\right)\underbrace{\iint\limits_{-\infty}^{\infty}{\left(2k_z\right)}^n\,\e^{-\mathrm{j}2\vec{k}\cdot\left(\vec{r}-\vec{r}_m\right)}\,\mathrm{d}{k}_x\,\mathrm{d}{k}_y}_{{F_{n}\left(\vec{r}-\vec{r}_m,  2k\right)}}\,\mathrm{d}^2\vec{r}_m\,,
	\end{equation}
	where \begin{equation}\label{k_case}
			k_z=\begin{cases}
				\sqrt{k^2-k_x^2-k_y^2} & \text{for~~$ k^2>k_x^2+k_y^2 $\,,}\\
				-\jm\sqrt{k_x^2+k_y^2-k^2} & \text{for~~$k^2<k_x^2+k_y^2$}
			\end{cases}\,
		\end{equation}
		denotes the dispersion relation of the wave number, which segments the planar plane-wave spectrum into propagating plane waves with $k^2>k_x^2\, +\,k_y^2$ and an evanescent region for $k^2<k_x^2 + k_y^2$. With the Weyl-identity in planar plane-wave representation as found in~(\ref{weyl_plane}) 
	and the derivative theorem of the Fourier transform, the improved spatial focusing operator $F_{n}\left(\vec{R}, 2k\right)$ for different orders $n$ of the spectral filter turns out to be
	\begin{equation}\label{comp_foc}
			F_{n}\left(\vec{R},2k\right)=\begin{cases}\frac{-2\uppi/4}{{\left(-\mathrm{j}\right)^n}}\frac{\partial^{n+1}}{ {\left(\partial R_z\right)}^{n+1}}\frac{\e^{-\mathrm{j}2kR}}{R}\,,R_z>0\\
				\frac{+2\uppi/4}{{\left(-\mathrm{j}\right)^n}}\frac{\partial^{n+1}}{ {\left(\partial R_z\right)}^{n+1}}\frac{\e^{+\mathrm{j}2kR}}{R}\,,R_z<0
			\end{cases}
		\end{equation}
		where $\vec{R}=\vec{r}-\vec{r}_m=R_x\uv{x}+R_y\uv{y}+R_z\uv{z}$ and  \mbox{$R=\sqrt{R_x^2+R_y^2+R_z^2}$} was utilized. Working out the derivatives for the cases \mbox{$n=0,1,2$} with the assumption $R_z<0$ as considered throughout this work results into 
	\begin{equation}\label{foc_op_n_0}
		F_{0}\left(\vec{R},2k\right) = \frac{\uppi}{2} \left(\frac{2\mathrm{j}kR_z}{R^2}-\frac{R_z}{R^3}\right)\e^{+\mathrm{j}2kR},~~~~~~~~~~~~~~~~~~~
	\end{equation}
	\begin{multline}\label{foc_op_n_1}F_{1}\left(\vec{R},2k\right)=	\frac{\mathrm{j}\uppi}{2}\left[\frac{4\mathrm{j}^2k^2R_z^2}{R^3}+\left(R_x^2+R_y^2-2R_z^2\right)\right.\\\left.\left(\frac{2\mathrm{j}k}{R^4}-\frac{1}{R^5}\right)\right]\e^{+\mathrm{j}2kR},
	\end{multline}
	\begin{multline}\label{foc_op_n_2}
		F_{2}\left(\vec{R},2k\right)=-\frac{\uppi}{2}\left[\frac{8\mathrm{j}^3{k}^3R_z^3}{R^4}+R_z\left(\frac{12\mathrm{j}^2{k}^2(R_x^2+R_y^2-R_z^2)}{R^5}+\right.\right.
		\\\left.\left.(3R_x^2+3R_y^2-2R_z^2)\left(-\frac{6\mathrm{j}k}{R^6}+\frac{3}{R^7}\right)\right)\right]\e^{+\mathrm{j}2kR}\,.
	\end{multline}
	If $R$ is sufficiently large, the expressions in (\ref{foc_op_n_0}), (\ref{foc_op_n_1}), and (\ref{foc_op_n_2}) can be reduced to its first terms only.
	Fig.~\ref{foc_examples} shows results for the spatial distribution of the magnitude correction factor $\abs{F_n(\vec{R},2k)}$ for $n=0$ and $n=2$ together with a planar scanning surface spanning an aperture size of 300\,mm by 300\,mm and for two different  distances of 30\,mm and 300\,mm from the aperture plane to the imaging domain. This plotting of the magnitude characteristics reveals how different parts in the aperture distribution, dependent on distance, contribute for the image creation of a single point scatterer located at the origin.
	The focusing operator with a filter order of $n=2$ shows obviously a larger directivity in the spatial domain than the one without filtering, i.e.,~\mbox{$n=0$}. As shown in Section~\ref{sim_setup}, a filter function with higher order of $n$ can effectively mitigate a larger amount of artifacts in the microwave images, but may also lead to a decrease of the image resolution.
	\begin{figure}[tbp]
		\centering
		\subfloat[]{\includegraphics[scale=0.84]{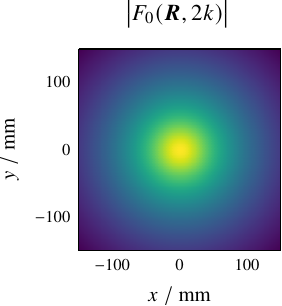}}\hfill
		\subfloat[]{\includegraphics[scale=0.84]{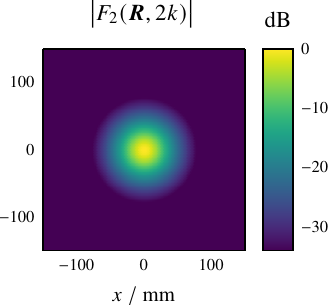}}\\
		\subfloat[]{\includegraphics[scale=0.84]{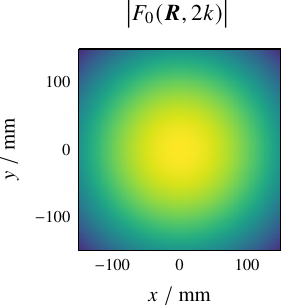}}\hfill
		\subfloat[]{\includegraphics[scale=0.84]{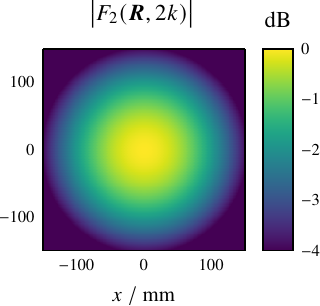}}
		\caption{Magnitude distributions of the focusing operator of orders $n=0$ and $n=2$ with different distances from the aperture plane to the imaging domain for $k=210\,{\mathrm{m}}^{-1}$. (a), (b)~30\,mm, (c), (d)~300\,mm.}
		\label{foc_examples}
	\end{figure}

	\section{Imaging results based on simulated data}\label{sim_setup}

   \subsection{Computational Complexity and Resolution Limits}
		Since the proposed inverse source algorithm as well the MLFSDA-$\omega$-$k$ approach both operate in the spatial frequency domain and are accelerated by a hierarchical multi-level scheme, the computation of the spectral representation of the scattering object for one single frequency has a complexity of $\mathcal{O}(N_xN_y\log(N_xN_y))$. Here, it is assumed that the number of spectral sample points along both lateral directions denoted by $N_x$ and $N_y$ is proportional to the number of measurements along the respective direction. In this work, the image creation for the $\omega$-$k$ based methods is performed by back-projecting the computed plane-wave spectra into different cut planes along the $z$-axis and coherently superimposing the contributions for different frequencies after using inverse 2D-FFTs or hierarchical disaggregation. Hence, if $N_z$ denotes the number of voxels in $z$-direction and if $N_f$ is the number of frequency samples, the computational complexity for the MLFSDA based $\omega$-$k$ method is $\mathcal{O}(N_fN_{z}N_xN_y\log(N_xN_y))$, which corresponds to the numerical cost employing the 2D-FFT based $\omega$-$k$ method. Compared to this, the complexity for the inverse source algorithm differs in an additional factor $N_{\mathrm{iter}}$ denoting the number of iterations in order to evaluate the repeated calls for the forward and adjoint operator. The image creation utilizing either the naive BPA without magnitude correction or the BPA employing the improved focusing operators as derived in this work show a complexity of $\mathcal{O}(N_fN_zN_x^2N_y^2)$. 
		For a first verification of the proposed imaging techniques, a fully polarimetric set of scattering data is synthetically generated using one single point scatterer located at the center of the coordinate system and Hertzian dipoles of different orientations, which serve as the Tx and Rx for this simulation scenario. The antenna elements are placed according to a Gauß-Legendre grid, which allows to employ a corresponding numerical quadrature rule for the BPA integral, which is, of course, irrelevant for the inverse source based imaging algorithm. The frequency is fixed to \mbox{110 GHz}, corresponding to a wavelength of $\lambda=2.7$\,mm. A total of \numprint{25600} observation points is irregularly distributed within an aperture size of 54.4\,mm by 54.4\,mm ($20\lambda$ by $20\lambda$) and the  distance from the observation aperture to the target is 13.5\,mm (5$\lambda$). Since the inverse source based imaging algorithm according to (\ref{gen_nf}), (\ref{disk_s}), and (\ref{ne_eq}) is more robust and,  hence, better suited to eliminate the influence of the probe patterns compared to the adjoint imaging techniques, Fig.~\ref{fig:ptscat_ideal}(a) and (b) show the probe-corrected reconstruction results for the ideal point scatterer without spectral filtering and considering the filter function $H_2(2\vec{k})=4k_z^2$.
	
		 As expected, both images from the inverse source approach exhibit circular symmetry in the reconstructed point spread function, and the utilization of spectral filtering reduces the sidelobe level quite significantly. Since the utilized sampling grid is relatively dense, the irregularly distributed measurement points are interpolated on a regular grid employing a fixed point Lagrange interpolation with eleven interpolation samples along each direction. Afterwards, the corresponding spatial image was computed very efficiently using the 2D-FFT-$\omega$-$k$ approach based on (\ref{gen_inf}) and as shown in Fig.~\ref{fig:ptscat_ideal}(c), where again the filter mask $H_0(2\vec{k})$ was employed. While the sidelobe level for this case along the $y$-axis is comparable to the inverse source solution without spectral filtering, the point spread function is also more distorted. The latter is due to the fact that for the 2D-FFT-$\omega$-$k$  method one co-polarized set of observations (using $x$-oriented dipoles) without probe-correction was utilized. Since the spectral inversion without additional filtering is mathematically equivalent to the BPA with the focusing operator of lowest order according to (\ref{foc_def}) and (\ref{foc_op_n_0}), the corresponding spatial image as given in Fig.~\ref{fig:ptscat_ideal}(d) shows good similarity to the image obtained by the 2D-FFT-$\omega$-$k$ algorithm.
			\begin{figure}
				\subfloat[]{\includegraphics[scale=0.84]{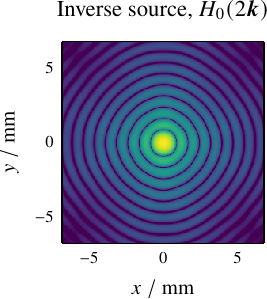}}\hfill
				\subfloat[]{\includegraphics[scale=0.84]{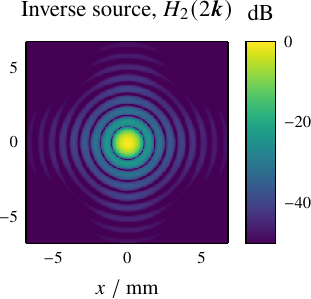}}\\
				\subfloat[]{\includegraphics[scale=0.84]{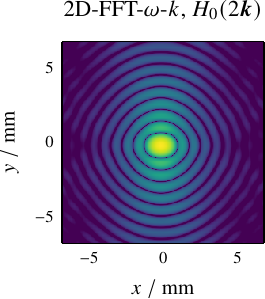}}\hfill
				\subfloat[]{\includegraphics[scale=0.84]{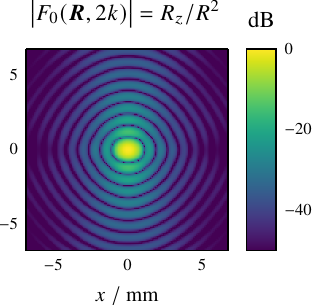}}
				\caption{Microwave images of a single point scatterer. (a), (b) Probe-corrected inverse source solution without filtering and employing the second order filter $H_2(2\vec{k})=4k_z^2$. Transforming one co-polarized set of observations using 2D-FFT-$\omega$-$k$ after interpolation to a regular grid (c) and the BPA with the proposed focusing operator $F_0(\vec{R},2k)$ (d).}
				\label{fig:ptscat_ideal}
			\end{figure}
		For a more rigorous analysis, cuts obtained for the different point spread functions along the $x$- and $y$-axis are given in Fig.~\ref{point_spread}(a) and (b), respectively. 
			\begin{figure}
			\subfloat[]{\includegraphics[scale=0.62]{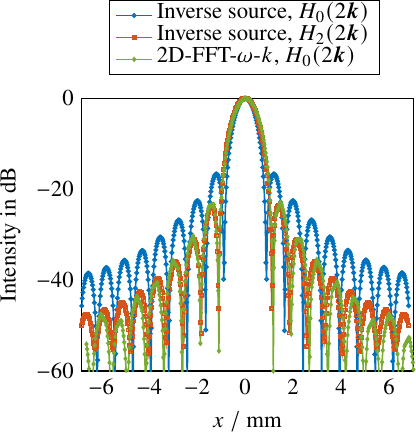}}\hfill
			\subfloat[]{\includegraphics[scale=0.62]{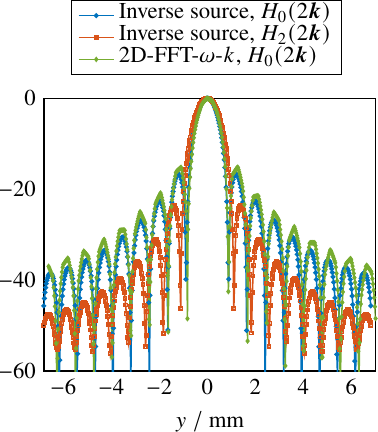}}
			\caption{Cuts of the reconstructed images as given in Fig.~\ref{fig:ptscat_ideal} along the $x$-axis (a) and along the $y$-axis (b).}
			\label{point_spread}
		\end{figure} 
	   Interestingly, the point spread function for the 2D-FFT-$\omega$-$k$ approach (which is basically identical to the case of the improved BPA with focusing operator $F_0(\vec{R},2k)$ and, therefore, dropped in Fig.~\ref{point_spread}) shows good similarity to the filtered inverse source solution along the $x$-axis, whereas along the $y$-axis it follows more closely the curve of the unfiltered inverse source reconstruction. A quantitative analysis for the different imaging algorithms including the sidelobe level $\mathrm{SLL}_{x,y}$ as well as the resolution limits denoted by $\delta_{x,y}$ along both lateral directions is given in Tab.~\ref{tab:pt_scat_res}.
	   \begin{table}
		\caption{Sidelobe level and obtained resolution limits for the images according to Fig.~\ref{fig:ptscat_ideal}.}
		\begin{tabular*}{87 mm}{c @{\extracolsep{\fill}} ccccc}
	    \toprule
	    method &	$\mathrm{SLL}_x$\,/\,dB  &  $\delta_x$\,/\,mm  &   $\mathrm{SLL}_y$\,/\,dB & $\delta_y$\,/\,mm \\
	    \cmidrule(lr){2-2}\cmidrule(lr){3-3}\cmidrule(lr){4-4}\cmidrule(lr){5-5}
	    \cmidrule(lr){1-1}
	    Inverse source, $H_0(2\vec{k})$  & -16.66       &  0.89      &  -16.60    & 0.89    \\
	    Inverse source, $H_2(2\vec{k})$   & -23.60     &  1.11    &  -23.60    &   1.11     \\
	    2D-FFT-$\omega$-$k$, $H_0(2\vec{k})$     & -22.89      &  1.17     &  -15.18      & 0.88   \\ 
	    \bottomrule
    \end{tabular*}
	\label{tab:pt_scat_res}
		\end{table}
		
	\subsection{Airplane Model of Point Scatterers}
	For the next simulations, the single point scatterer is replaced by a model of 51 ideal (isotropic) point scatterers, which are located such that they are forming
		the outline of an airplane. Regarding the Tx and Rx positions two irregular sampling grids are investigated, where the first one is identical to the Gauß-Legendre grid utilized in the previous example. The second aperture distribution to be considered is resulting by additionally varying the positions of the Gauß-Legendre grid in all directions. A graphical representation of the scattering scenarios utilizing both grids is shown in Fig.~\ref{plane_sim_setup}(a) and (b), respectively. Additionally, we want to emphasize that all considered reconstruction algorithms were implemented in MATLAB~\cite{MATLABversion:9.10.0.} and all images were computed for 256 by 256 pixels using eight threads on an Intel Xeon processor E5-1630.
     
	\begin{figure}
		\centering
		\subfloat[]{\includegraphics[scale = 0.8]{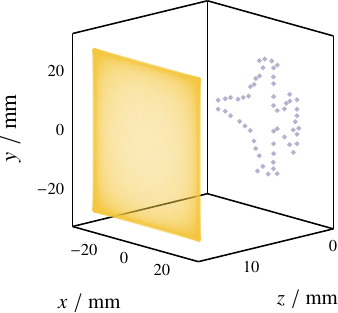}}\hfill
		\subfloat[]{\includegraphics[scale = 0.8]{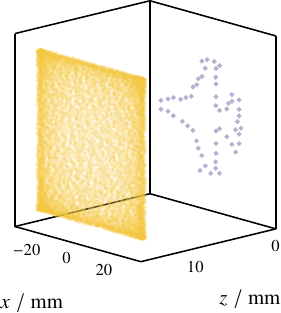}}
		\caption{Visualization of the imaging setups with the airplane model consisting of point scatterers with observation sample locations according to a Gauß-Legendre grid~(a), which is additionally varied at random in all directions~(b).}
		\label{plane_sim_setup}
	\end{figure}
	As a first validation of our theoretical considerations, the result of the inverse source reconstruction based on (\ref{gen_nf}), (\ref{disk_s}), and (\ref{ne_eq}) with filter function $H_0(2\vec{k})=1$ is compared to the image generated via the spatial BPA employing the corresponding magnitude correction factor $R_z/R^2$ according to (\ref{foc_def}) and (\ref{foc_op_n_0}). As expected and as can be seen in Fig.~\ref{Flugzeug_GL_ref}(a) and (b), the reconstruction results for both approaches show excellent agreement in terms of their focusing capabilities. The image for the standard BPA according to (\ref{naive_bpa}) and as shown in Fig.~\ref{Flugzeug_GL_ref}(d), which does not include the correct focusing operator, is heavily distorted. In order to utilize the 2D-FFT-$\omega$-$k$ method, again a Lagrange interpolation onto a regular grid is performed prior to the image creation according to (\ref{gen_inf}). Despite overall good focusing, the generated image depicted in Fig.~\ref{Flugzeug_GL_ref}(c) shows minor artifacts at the edges of the image domain resulting from interpolation errors.\begin{figure}
		\centering
		\subfloat[]{\includegraphics[scale=0.82]{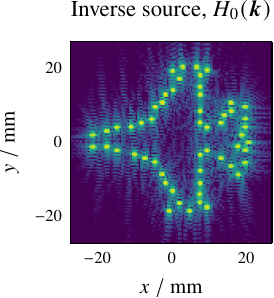}}\hfill
		\subfloat[]{\includegraphics[scale=0.82]{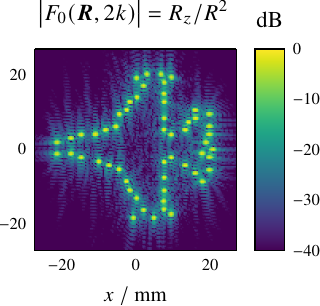}}\\
		\subfloat[]{\includegraphics[scale=0.82]{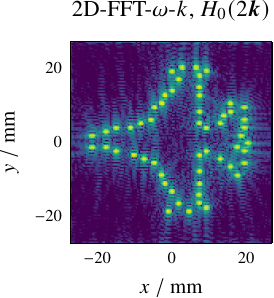}}\hfill
		\subfloat[]{\includegraphics[scale=0.82]{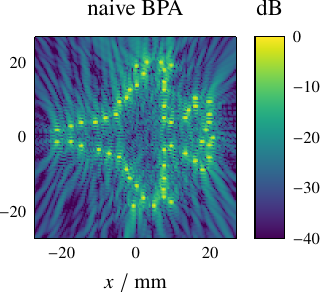}}
		\caption{Unfiltered imaging results for the synthetically generated scattering data according to Fig.~\ref{plane_sim_setup}(a) employing a Gauß-Legendre grid. (a) Inverse source reconstruction, (b), (d) spatial BPAs, (c) 2D-FFT-$\omega$-$k$.} 
		\label{Flugzeug_GL_ref}
	\end{figure}
	 Considered next is the effect of the filter function $H_{2}(2\vec{k})=4k^2_z$ as related to the improved BPA with spatial focusing operator $F_{n}(\vec{R},2k)$ of order $n=2$. The corresponding reconstruction results of the inverse source based approach as given in Fig.~\ref{Flugzeug_GL}(a) and of the MLFSDA based $\omega$-$k$-algorithm according to (\ref{gen_inf}) and as found in Fig.~\ref{Flugzeug_GL}(b) are here of similarly good quality, which is mostly due to the Gauß-Legendre grid of the observation points and the use of the correct quadrature rule. Compared to the results without filter function in Fig.~\ref{Flugzeug_GL_ref}(a) and (b), the filtered images show a considerably reduced amount of artifacts for the cost of slightly larger point scatterer spots.
	To further demonstrate the performance of the improved focusing operators, we consider another operator as found in Watanabe et al.~\cite{Watanabe.Mar.2022}. In this work, the goal was to derive a general expression of the focusing operator for arbitrary scanning surfaces, mainly based on the works in~\cite{Osipov.Sep.2013,Broquetas.May1998}. The definition of this focusing operator (see, e.g., (7) in~\cite{Watanabe.Mar.2022}) is identical to our definition given in (\ref{foc_def}), but its derivation is based on an operator inversion in the spatial domain. The corresponding focusing operator for the special case of a planar scanning surface $z=z_m$ is~\cite{Watanabe.Mar.2022}
	\begin{equation}\label{WAT_foc}
		F\left(\vec{R},2k\right)=\frac{R_z}{R}\e^{\,\mathrm{j}2kR}\,.
	\end{equation}
	The imaging result obtained with this operator in combination with (\ref{foc_def}) is depicted in Fig.~\ref{Flugzeug_GL}(c) and shows an even larger amount of artifacts than the results in Fig.~\ref{Flugzeug_GL_ref}(a) and (b). By comparison, the reconstruction result of the improved BPA with the second order focusing operator obtained by (\ref{foc_def}) and (\ref{foc_op_n_2}), and as depicted in Fig.~\ref{Flugzeug_GL}(d) shows not only again good image quality but also exhibits good similarity to the filtered inverse source solution and the filtered MLFSDA based $\omega$-$k$ method.
	\begin{figure}
	\subfloat[]{\includegraphics[scale=0.82]{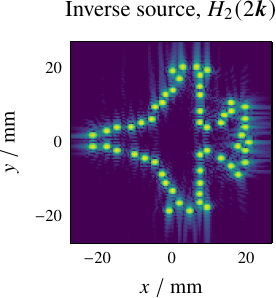}}\hfill
	\subfloat[]{\includegraphics[scale=0.82]{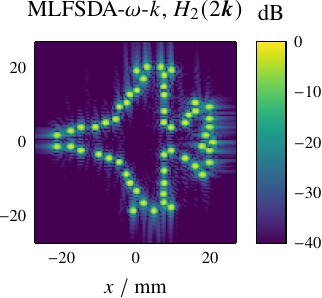}}\\
	\subfloat[]{\includegraphics[scale=0.82]{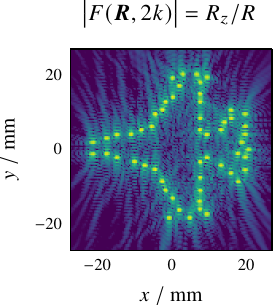}}\hfill
	\subfloat[]{\includegraphics[scale=0.82]{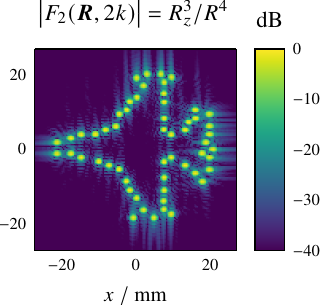}}
	\caption{Filtered imaging results for the synthetically generated scattering data according to Fig.~\ref{plane_sim_setup}(a) employing a Gauß-Legendre grid. (a) Inverse source solution, (b) MLFSDA based $\omega$-$k$ algorithm, (c), (d) spatial BPAs.} 
	\label{Flugzeug_GL}
\end{figure}
 In order to demonstrate the better flexibility and robustness of the inverse source based imaging algorithm, the scattering scenario as depicted in~Fig.~\ref{plane_sim_setup}(b) is considered next. Here, the observation locations compared to the previously utilized Gauß-Legendre grid are
 varied at random in all directions according to a uniform distribution within the interval \mbox{[-\,0.3\,mm, 0.3\,mm]}. The corresponding reconstruction results for this scenario are given in Fig.~\ref{Flugzeug_GL_var}(a)-(d). 
  \begin{figure}
 	\subfloat[]{\includegraphics[scale=0.82]{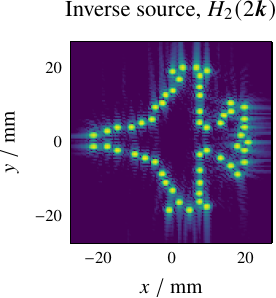}}\hfill
 	\subfloat[]{\includegraphics[scale=0.82]{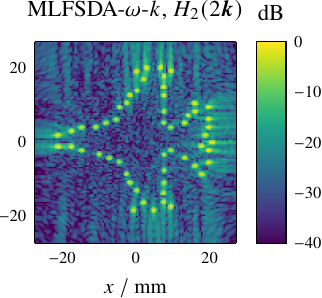}}\\
 	\subfloat[]{\includegraphics[scale=0.82]{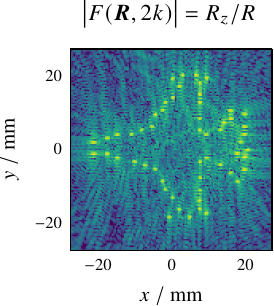}}\hfill
 	\subfloat[]{\includegraphics[scale=0.82]{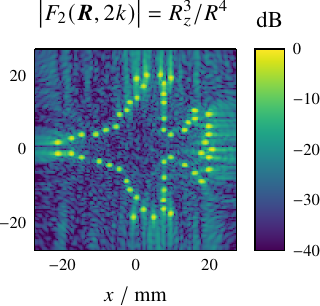}}
 	\caption{Filtered imaging results for the synthetically generated scattering data according to Fig.~\ref{plane_sim_setup}(b)  employing a Gauß-Legendre grid, which is additionally varied in all directions at random. (a) Inverse source solution, (b) MLFSDA based $\omega$-$k$ algorithm, (c), (d) spatial BPAs.}
 	\label{Flugzeug_GL_var}
 \end{figure}
The inverse source based imaging algorithm leads again to a well-focused spatial image, which is basically identical to the reconstruction result from the previous case. All the other methods generate a significant amount of artifacts, which is mainly caused by falsely evaluating the underlying integrals needed for the image generation. As depicted in Fig.~\ref{first_sim_conv}, a relative stopping criterion of $\varepsilon_{\mathrm{rel}}=0.99$ for terminating the GMRES-solver within the iterative inverse source based imaging algorithm was achieved after 91 iterations for case 1 (no variation in range direction) and after 50 iterations for case 2 (random variation in all directions). Thus, the inverse source based method clearly offers advantages when dealing with irregular sampling distributions.
	\begin{figure} 
		\centering
		\includegraphics[scale = 0.8]{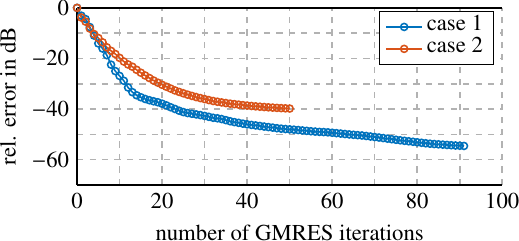}
		\caption{Convergence behavior of the inverse source based approach employing the GMRES-solver. Case 1: planar setup with a Gauß-Legendre grid. Case 2: random variation in all directions.}
		\label{first_sim_conv} 
	\end{figure}

    With respect to a quantitative analysis, we utilize the power ratio $\eta=P_{\mathrm{O}}/P_{\mathrm{I}}$ as introduced in~\cite{Wang.Jan.2020}, where $P_{\mathrm{O}}$ denotes the accumulated intensities outside a region of interest, while $P_{\mathrm{I}}$ represents the accumulated image intensities inside this region. Similar to using image entropy as a metric to assess the image quality~\cite{Zhang.Jul.2015}, a better image is, therefore, characterized by a smaller value for the power ratio. With respect to the synthetically generated scattering data using the airplane model, the region of interest can be conveniently defined and evaluated by placing small boxes around the individual point scatterers. The image intensities within these boxes are summed up to $P_{\mathrm{I}}$, whereas all other image intensities elsewhere contribute to $P_{\mathrm{O}}$. Additionally, the spatial extents of these boxes correspond to the resolution limits as found for the unfiltered inverse source solution given in Tab.~\ref{tab:pt_scat_res}. The quantitative assessment for the two considered scattering scenarios is given in Tab.~\ref{Tab_num_GL} and Tab.~\ref{Tab_num_GL_var}, respectively. As indicated by the reconstructed images, the obtained values for the power ratio and the image entropy utilizing the MLFSDA-$\omega$-$k$ with the filter mask $H_2(2\vec{k})$ agree very well with the improved BPA employing the focusing operator with magnitude $\abs{F_2(\vec{R},2k)}=R^3_z/R^4$. Tab.~\ref{Tab_num_GL} and Tab.~\ref{Tab_num_GL_var} also include the computation times, where it is obvious that the iterative inverse source algorithm is the most time-expensive reconstruction scheme. However, the filtered inverse source solution also shows for both scattering scenarios the best results with respect to power ratio and entropy.\begin{table}
  	\caption{Power ratio,  image entropy, and computation time utilizing the airplane model and the Gauß-Legendre grid according to Fig.~\ref{plane_sim_setup}(a).}
  	\begin{tabular*}{87 mm}{c @{\extracolsep{\fill}} cccc}
  		\toprule
  		method &	~~power ratio $\eta$ in \% &  entropy  & time in s\\
  		\cmidrule(lr){2-2}\cmidrule(lr){3-3}\cmidrule(lr){4-4}
  		\cmidrule(lr){1-1}
  		Inverse source, $H_0(2\vec{k})$~ &  10.18      &  8.13         &  728.70    \\
  		Inverse source, $H_2(2\vec{k})$~ &  3.35       &  8.07         &  714.96   \\
  		MLFSDA-$\omega$-$k,H_2(2\vec{k})$&  3.63       &  8.09         &  13.82   \\
  		$\abs{F_0(\vec{R},2k)}=R_z/R^2$  &  10.41      &  8.16         &  99   \\
  		$\abs{F_2(\vec{R},2k)}=R^3_z/R^4$&  3.63       &  8.09         &  390   \\ 
  		2D-FFT-$\omega$-$k$,  $H_0(2\vec{k})$                  &  9.86       &  8.16         &  22   \\
  		naive BPA                        &  55.54      &  9.2          &  90   \\
  	  	$\abs{F(\vec{R},2k)}=R_z/R$      &  22.66      &  8.49         &  98   \\
  		\bottomrule
  	\end{tabular*}
  	\label{Tab_num_GL}
  \end{table}\begin{table}[h!]
	\caption{Power ratio,  image entropy, and computation time utilizing the airplane model and the Gauß-Legendre grid with additional variations according to Fig.~\ref{plane_sim_setup}(b).}
\begin{tabular*}{87 mm}{c @{\extracolsep{\fill}} cccc}
	\toprule
	method &	~~power ratio $\eta$ in \% &  entropy  & time in s\\
	\cmidrule(lr){2-2}\cmidrule(lr){3-3}\cmidrule(lr){4-4}
	\cmidrule(lr){1-1}
	Inverse source, $H_2(2\vec{k})$~ & 3.76       &  8.05      &  353.24   \\
	MLFSDA-$\omega$-$k,H_2(2\vec{k})$& 30.12      &  8.86      &  14.08   \\
	$\abs{F_2(\vec{R},2k)}=R^3_z/R^4$& 30.23      &  8.86      &  386   \\ 
	$\abs{F(\vec{R},2k)}=R_z/R$      & 95.13      &  9.61      &  102   \\
	\bottomrule
\end{tabular*}
\label{Tab_num_GL_var}
\end{table}

	\subsection{Full-Wave Simulation Utilizing Coarse Sampling Grids}
	The scattering simulations for the next scenario were performed in CST MWS~\cite{CSTComputerSimulationTechnology.} at a fixed frequency of 40\,GHz utilizing again Hertzian dipoles as transmitting and receiving antennas. The target consists of the letters HFT and is assumed perfectly electrically conducting. The uniform observation sampling grid utilizes 80 by 80 sampling points spanning an aperture size of 200\,mm by 200\,mm at a distance of 112.5\,mm from the target. Simulations were conducted for different orientations of the transmitting and receiving antennas in order to obtain a fully-polarimetric set of scattering data. In order to showcase that the inverse source based approach performs well even for undersampled data, 100\%, 90\%, 80\%, and 70\%  of the originally available data are chosen at random. For a fair comparison between the inverse source based imaging algorithm and the MLFSDA based $\omega$-$k$-algorithm, a Voronoi decomposition with respect to the observation points was performed~\cite{Wang.Jan.2020}. As such, the areas of all cells containing an observation point served as the quadrature weights, which were utilized in the computation of the plane-wave representation of the observation data in the MLFSDA based $\omega$-$k$-approach. For the observation points, which are at the edges of the aperture plane, the maximum value of its neighboring cells was used as a weighting factor. 
\begin{figure*}[htb]
		\centering
		\subfloat[]{\includegraphics[scale=0.82]{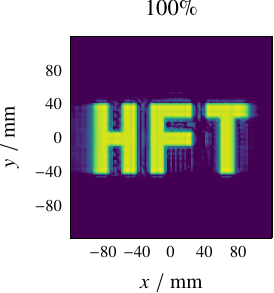}}\hfill
		\subfloat[]{\includegraphics[scale=0.82]{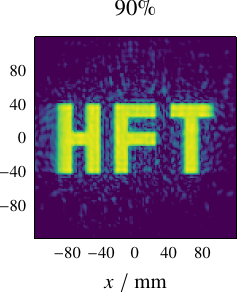}}\hfill
		\subfloat[]{\includegraphics[scale=0.82]{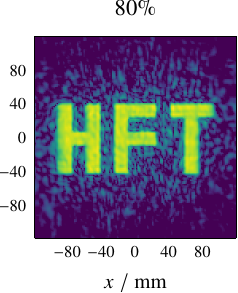}}\hfill
		\subfloat[]{\includegraphics[scale=0.82]{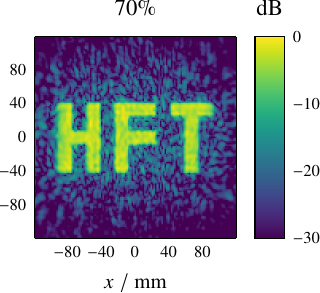}}\\
		\subfloat[]{\includegraphics[scale=0.82]{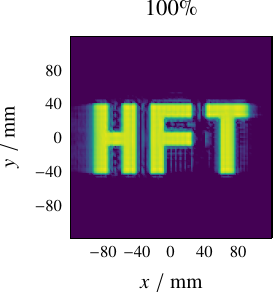}}\hfill
		\subfloat[]{\includegraphics[scale=0.82]{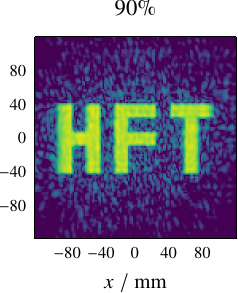}}\hfill
		\subfloat[]{\includegraphics[scale=0.82]{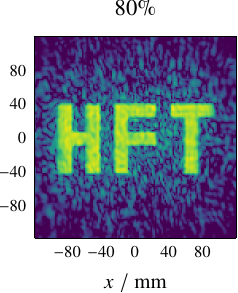}}\hfill
		\subfloat[]{\includegraphics[scale=0.82]{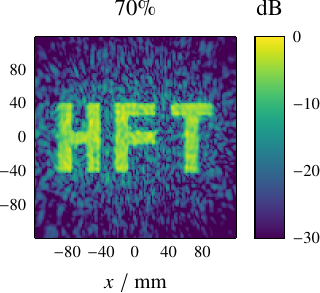}}
		\caption{Co-polarized scattering component $\left|s_{\varphi\varphi}\right|$ for the HFT logo employing the inverse source based approach (a)-(d), and the MLFSDA based $\omega$-$k$-algorithm (e)-(h), when utilizing different fractions of randomly chosen observation points.} 
		\label{HFT_logo_rek_IS_MF}
	\end{figure*}\begin{figure}
	\centering
	\includegraphics[scale = 0.75]{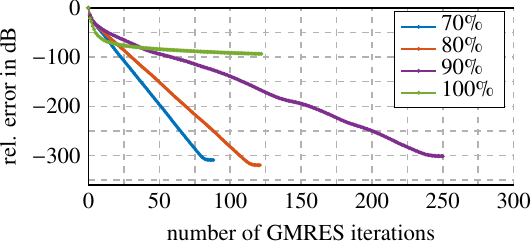}
	\caption{Convergence behavior of the inverse source based approach employing the GMRES-solver for the four different considered fractions of observation points.}
	\label{hft_series_conv} 	
\end{figure} 

The corresponding images, which are all based on employing the filter function $H_{2}(2\vec{k})=4k^2_z$, are depicted in Fig.~\ref{HFT_logo_rek_IS_MF}(a)-(d) for the inverse source based solution and in Fig.~\ref{HFT_logo_rek_IS_MF}(e)-(h) for the MLFSDA based $\omega$-$k$ approach. From a visual inspection, it already becomes obvious that the inverse source based method significantly performs better in terms of overall image quality. Evaluating the power ratio  as well as the image entropy for all the different considered cases as summarized in Tab.~\ref{Tab_pts}, confirms these findings. As expected, the largest visually noticeably differences appear in the case when only 70\% of the originally available observation samples are utilized. Important to note is that in this case the maximum distance between two neighboring sampling points is approximately 3.6\,$\mathrm{mm}$, which is significantly larger than the required sampling distance of 2.2\,mm to satisfy the Nyquist criterion~\cite{LopezSanchez.May2000}. Thus, the inverse source based approach also strongly mitigates aliasing artifacts. While the computation times for both methods can be also found in Tab.~\ref{Tab_pts}, the convergence rates for the iterative solver are depicted in Fig.~\ref{hft_series_conv}. The required number of solver iterations to achieve a desired relative error level of $\varepsilon_{\mathrm{rel}}=0.99$ is the largest when 90\% of the observation samples are utilized. As can be seen Fig.~\ref{HFT_logo_rek_IS_MF}(b) and (f), respectively, and as shown in Tab.~\ref{Tab_pts}, the inverse source based solution leads to a significantly better focused image with a strongly reduced amount of aliasing artifacts in this case as well.
\begin{table}
	\caption{Numerical evaluation and computation times for the HFT logo utilizing different amount of sampling points.}
	\begin{tabular*}{87 mm}{c @{\extracolsep{-2pt}} cccc}
		\toprule
		parameter &	~~~sample points & Inverse source &  MLFSDA-$\omega$-$k$\\
		\cmidrule(l){2-2}\cmidrule(lr){3-3}\cmidrule(lr){4-4}
		\cmidrule(lr){1-1}\cmidrule(lr){2-2}
		\multirow{4}{*}{~~~power ratio in \,\%~~~} & ~~~70\% & 7.24 &14.1 \\
		& ~~~80\% & 4.62 & 9.69\\
		& ~~~90\% & 2.08 & 4.74 \\
		& ~~~100\% & 1.04 & 1.04\\
		\cmidrule(lr){1-1}\cmidrule(lr){2-2}\cmidrule(lr){4-4}\cmidrule(lr){3-3}
		\multirow{4}{*}{image entropy}  & ~~~70\% &   9.35   &  9.54   \\
		& ~~~80\% &  9.25   &    9.42    \\
		& ~~~90\% &   9.14   &   9.25      \\
		& ~~~100\% &   9.09   &   9.09   \\
		\cmidrule(lr){1-1}\cmidrule(lr){2-2}\cmidrule(lr){4-4}\cmidrule(lr){3-3}
		\multirow{4}{*}{time\,/\,s}  & ~~~70\% & \numprint{3850}   &  \numprint{1571}    \\
		& ~~~80\% &     \numprint{5287}    &     \numprint{1527}    \\
		& ~~~90\% &   \numprint{10870}   &       \numprint{1534}   \\
		& ~~~100\% &    \numprint{5506}   &   \numprint{1527}    \\
		\bottomrule
	\end{tabular*}
	\label{Tab_pts}
\end{table}

	\section{Measurement results}
	\label{meas}
	\subsection{Blister Package}
	Next, the performances of the inverse source based imaging algorithm and of the MLFSDA based $\omega$-$k$-algorithm are demonstrated for real measurement data\footnote{The measurement data was provided by the innovation department of Balluff GmbH, Neuhausen a.d.F., Germany.}. The measurement data has been acquired by the ultra wideband single channel transceiver TRA\_120\_031 by Silicon Radar GmbH with on-chip antennas. The test object is a partially emptied blister package, which is placed approximately 8\,cm away from the radar frontend as can be seen in Fig.~\ref{sirad_meas}. 
	\begin{figure}
		\centering 
		\includegraphics[width=\textwidth/2]{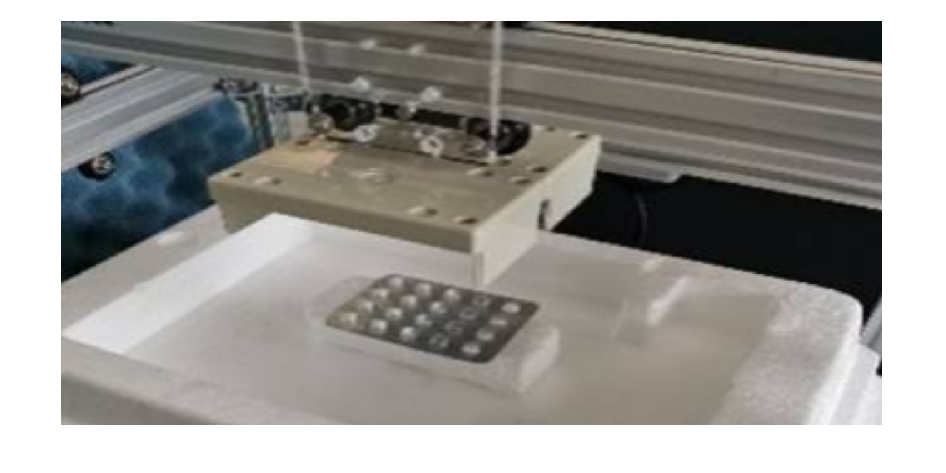}
		\caption{Data acquisition for a blister package utilizing a quasi-monostatic radar in an anechoic chamber.}
		\label{sirad_meas}
	\end{figure}
	Since the utilized transceiver is enclosed by a QFN package with lateral sizes of $5\,\mathrm{mm}$ by $5\,\mathrm{mm}$, the setup can be considered as quasi-monostatic. In this experiment, $M=161\times 161=25\,291$ regularly distributed measurement samples were acquired for each of the 128 discrete frequency samples, which cover a frequency range from 120\,GHz to 139.86\,GHz. The lateral dimensions of the synthesized virtual aperture are $20\,\mathrm{cm}$ by $20\,\mathrm{cm}$.
	The magnitude and phase distributions of the measurements performed at a frequency of 120\,GHz are depicted in Fig.~\ref{Sirad_amp_phase}(a) and (b) and give only a rough idea about the shape and position of the target.
	\begin{figure}
		\centering
		\subfloat[]{\includegraphics[scale=0.85]{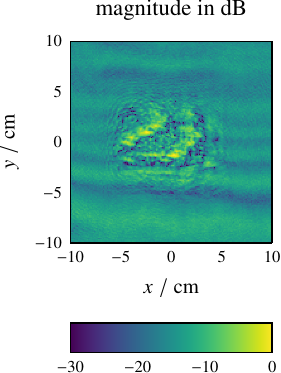}}\hfill
		\subfloat[]{\includegraphics[scale=0.85]{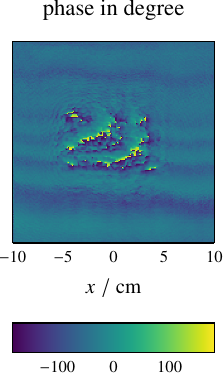}}
		\caption{(a) Magnitude and (b) phase distribution for the measured (co-polarized) scattering
			component. This measurement was performed at a frequency of $120\,\mathrm{GHz}$.}
		\label{Sirad_amp_phase}
	\end{figure}
	\begin{figure}
		\centering
		\subfloat[]{\includegraphics[scale=0.62]{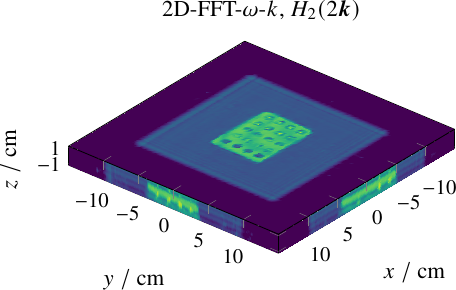}}\hfill
		\subfloat[]{\includegraphics[scale=0.65]{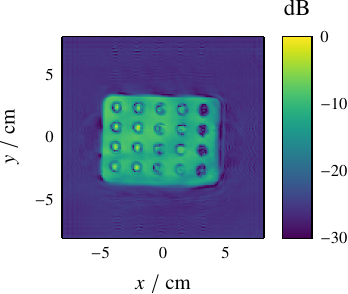}}\\
		\subfloat[]{\includegraphics[scale=0.62]{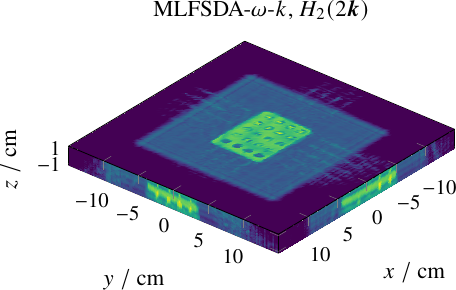}}\hfill
		\subfloat[]{\includegraphics[scale=0.65]{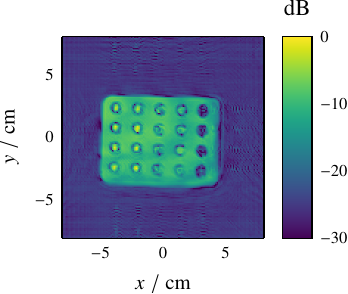}}\\
		\subfloat[]{\includegraphics[scale=0.62]{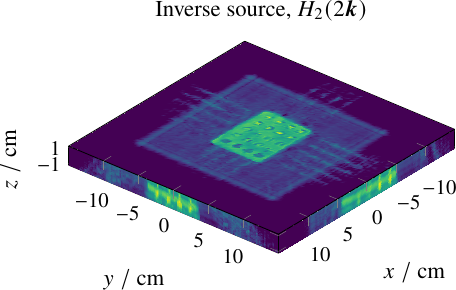}}\hfill
		\subfloat[]{\includegraphics[scale=0.65]{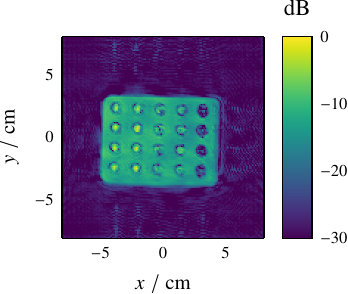}}
		\caption{Reconstruction results for the blister package as measured in Fig.~\ref{sirad_meas} using 2D-FFT-$\omega$-$k$ (a), (b), MLFSDA-$\omega$-$k$-algorithm (c), (d) and the inverse source solution (e), (f). Maximum projection of the computed intensity values to the faces of the reconstruction volume are shown in (a), (c) and (e), while (b), (d) and (f) show a planar cut through the reconstruction volume at $z=0$.} 
		\label{Blister_rek}
	\end{figure}The reconstructed spatial images, which are depicted in Fig.~\ref{Blister_rek}(a)-(f) have been created utilizing the 2D-FFT based $\omega$-$k$ method, the MLFSDA based $\omega$-$k$-algorithm, and the presented inverse source based method. For all three image reconstruction methods, the filter function
 $H_2(2\vec{k})=4k_z^2$ was applied to the obtained spectral representation of the measurement data  prior to the image creation. For a fair comparison of the different methods, the observation points are zero-padded utilizing the 2D-FFT-$\omega$-$k$ approach, which enables image reconstruction in the same domain as for the other two approaches. More specifically, the imaging domain has spatial extents of 30 cm by 30 cm by 2.4 cm and consists of 256 by 256 by 32 voxels. Overall, the reconstructed 3D image for the 2D-FFT based $\omega$-$k$ approach, which is depicted in Fig.~\ref{Blister_rek}(a) is very smooth and shows a good focusing of the blister package. The 3D spatial image for the MLFSDA based $\omega$-$k$-algorithm given in Fig.~\ref{Blister_rek}(c) is comparable in terms of image quality but also mitigated by artifacts around the blister package due to parasitic scattering contributions from the measurement setup and noise. These artifacts are suppressed by about 3.5\,dB when utilizing the iterative inverse source based imaging method as depicted in Fig.~\ref{Blister_rek}(e). Consequently, the dynamic range of the spatial image is significantly increased. This can be observed even more clearly by comparing slices through the imaging domain at $z=0$, which are given in Fig.~\ref{Blister_rek}(d) for the MLFSDA based $\omega$-$k$-algorithm and in Fig.~\ref{Blister_rek}(f) for the inverse source based solution. The number of unknowns in the inverse source solver was \numprint{448601} for the smallest frequency and \numprint{659344} for the largest frequency and the stopping criterion of $\varepsilon_{\mathrm{rel}}=0.99$ for the inverse source solution was achieved with below 40 iterations for all 128 frequencies. For a quantitative analysis,  the region for the computation of $P_{\mathrm{I}}$ is defined as a cuboid with dimension 10 cm by 8 cm by 2 cm. Based on this, the power ratio was computed and the corresponding values are given together with the image entropy and the computation time in Tab.~\ref{tab:eval_sirad}, where it is clearly found that the inverse source solution performs the best with respect to overall image accuracy.
\begin{table}
		\caption{Power ratio,  image entropy, and computation time utilizing the measurements for the partially emptied blister package.}
		\begin{tabular*}{87 mm}{c @{\extracolsep{\fill}} cccc}
			\toprule
			method &	~~power ratio $\eta$ in \% &  entropy  & time in s\\
			\cmidrule(lr){2-2}\cmidrule(lr){3-3}\cmidrule(lr){4-4}
			\cmidrule(lr){1-1}
			Inverse source, $H_2(2\vec{k})$~  & 37.69      &  12.46      & \numprint{225768}\\
			MLFSDA-$\omega$-$k,H_2(2\vec{k})$ & 51.69      &  12.67      &  3929.60         \\
		    2D-FFT-$\omega$-$k$, $H_2(2\vec{k})$     & 51.75      &  12.65      &  54.05           \\ 
			\bottomrule
		\end{tabular*}
		\label{tab:eval_sirad}
\end{table}

	\subsection{Metallic Plate Containing Different Evaluation Patterns}
	To further evaluate the performance of the inverse source imaging algorithm, a co-polarized set of scattering data was collected utilizing a Rohde \& Schwarz QAR50~\cite{Brinkmann.Sep.2023}, which possesses a multistatic arrangement of 94 transmitters and 94 receivers as depicted in Fig.~\ref{fig:setupandobjects}(a). The measurement utilizes 128 linearly distributed frequencies ranging from 76\,GHz to 81\,GHz. The considered scattering object as shown in Fig.~\ref{fig:setupandobjects}(b) is a metallic plate containing visualization patterns of different size and shape including circular and rectangular cutouts as well as a Siemens star with a radius of 30\,mm. The distance from the metallic plate to the aperture plane is approximately 20\,cm. Since the focus in this work is on monostatic inverse source based and adjoint imaging algorithms, the multistatic antenna configuration is converted into an equivalent monostatic arrangement employing the half-angle principle~\cite{Buddendick.2009}, which leads to an irregularly distributed quasi-planar aperture distribution as represented by the orange dots in Fig.~\ref{fig:setupandobjects}(a).
	\begin{figure}
		\centering%
		\subfloat[\label{fig:cluster}]{\includegraphics[scale=0.9]{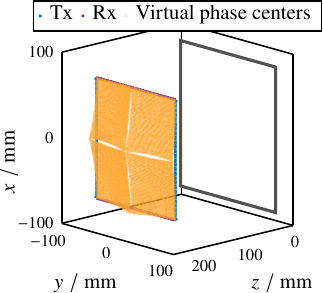}}\hfill%
		\subfloat[\label{fig:DUTs}]{\includegraphics[scale = 0.05]{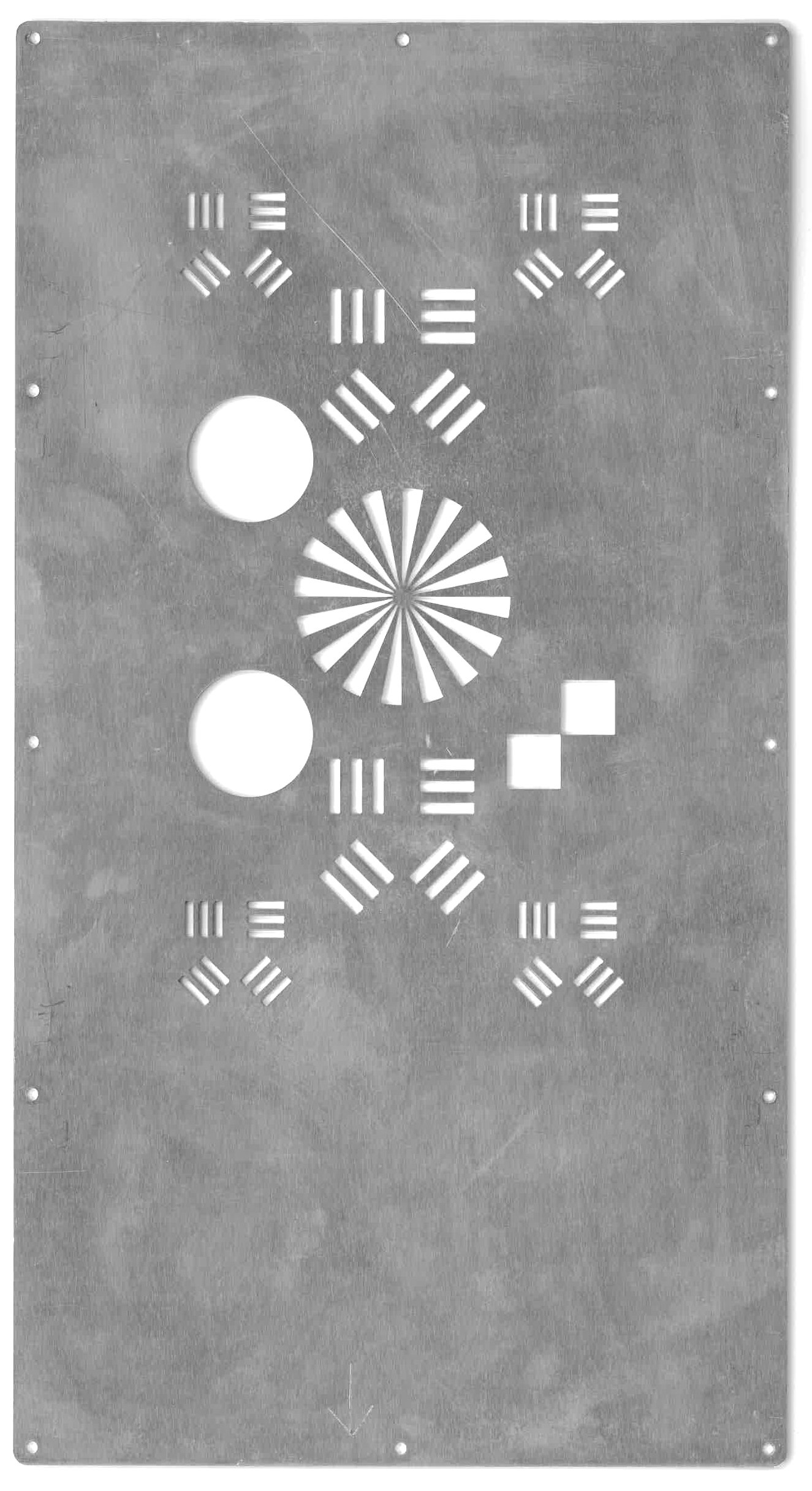}}
		\caption{Antenna arrangement of the Rohde \& Schwarz QAR50 (a). The orange dots represent the equivalent monostatic array, while the grey outline shows the border of the imaging domain. Picture of the evaluation object (b).}
		\label{fig:setupandobjects}
	\end{figure}
	
	The spatial images computed in the plane $z=0$ for the inverse source based reconstruction algorithm and the MLFSDA based $\omega$-$k$-algorithm are given in Fig.~\ref{reks_RUS_mf}(a) and (b), respectively. Prior to the image generation, the computed plane-wave spectra are multiplied by the second order filter function $H_2(2\vec{k})=4k_z^2$. The total amount of solver iterations required to achieve once again a relative stopping criterion of $\varepsilon_{\mathrm{rel}}=0.99$ for all frequencies was between 40 and 50. For further comparison, the microwave images originating from the improved BPA with focusing operator of second order $F_2(\vec{R},2k)$ as well as the naive monostatic BPA are given in~Fig.~\ref{reks_RUS_mf}(c) and (d), respectively. Closely inspecting the different spatial images reveals that the inverse source  based imaging method, which is apparently more effective in dealing with irregular sampling points, leads to a significantly better focusing of the Siemens star in the central region of the reconstruction domain compared to the other three adjoint imaging approaches.
	\begin{figure}
		\centering
		\subfloat[]{\includegraphics[scale=0.82]{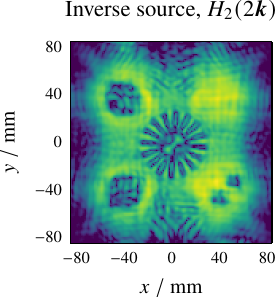}}\hfill
		\subfloat[]{\includegraphics[scale=0.82]{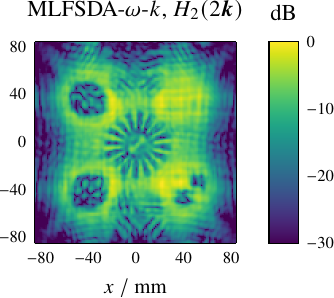}}\\
		\subfloat[]{\includegraphics[scale=0.82]{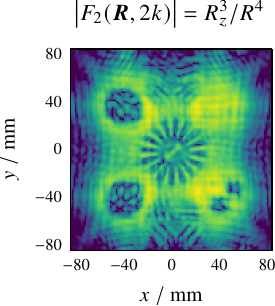}}\hfill
		\subfloat[]{\includegraphics[scale=0.82]{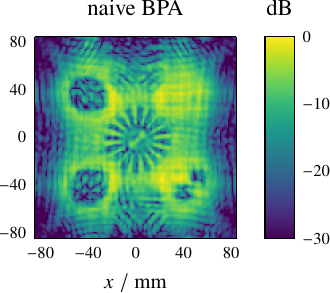}}\\
		\subfloat[]{\includegraphics[scale=0.82]{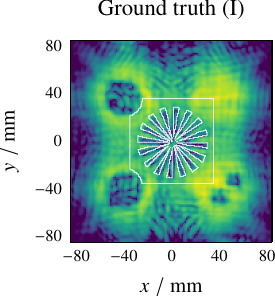}}\hfill
		\subfloat[]{\includegraphics[scale=0.82]{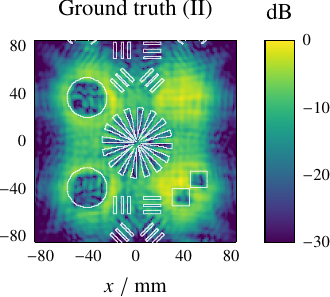}}
		\caption{(a), (b) Filtered reconstruction results for the inverse source based imaging algorithm and the MLFSDA based $\omega$-$k$-algorithm, (c), (d) images for the spatial BPAs, (e), (f) utilizing different ground truths and the inverse source solution as depicted in (a) to compute the power ratio and image entropy.}
		\label{reks_RUS_mf}
	\end{figure}
   In order to quantitatively assess the reconstruction results, the ground truth  of the verification plate is utilized. More specifically, two segmentation masks as shown by their white outlines in Fig.~\ref{reks_RUS_mf}(e) and (f) are employed. The first one has a spatial extent of 70\,mm by 70\,mm centered at the origin, which, thus, captures only the Siemens star, while the second one corresponds to the ground truth of the metallic plate in the complete imaging domain. The results for a subsequent numerical evaluation for the two cases as well as the reconstruction times are given in Tab.~\ref{rus:pt_scat_res}. It can be stated that for the first ground truth denoted by GT (I), the inverse source solution yields by far the most accurate results, which is in good agreement to the spatial images according to Fig.~\ref{reks_RUS_mf}(a)-(d). For the second ground truth as indicated by GT (II), the inverse source solution shows a slightly larger power ratio compared to the other methods, while the image entropy is comparable. Since the multistatic-to-monostatic conversion is, however, less accurate further away from the center, this finding should be considered with care. It is also worth noting that due to the utilization of multiple frequencies the naive monostatic BPA is only slightly worse compared to the result from the MLFSDA based $\omega$-$k$ method and the improved BPA, which, as expected, show identical focusing capabilities.
\begin{table}
	\centering
	\caption{Numerical evaluation of the different reconstruction results based on the measurements provided by the Rohde \& Schwarz QAR50.}
		\begin{tabular*}{87 mm}{c@{\extracolsep{-2pt}}ccccc}
	\toprule
	\multirow{2}{*}{method} &		\multicolumn{2}{c}{$\eta$ in \%} &  	\multicolumn{2}{c}{entropy} &\multirow{2}{*}{time\,/\,s} \\
	  &   ~~GT (I)   &    GT (II) &    GT (I)   &  GT (II) & \\
	    \cmidrule(lr){1-1}\cmidrule(lr){2-2}\cmidrule(lr){3-3}\cmidrule(lr){4-4}\cmidrule(lr){5-5}\cmidrule(lr){6-6}
	  Inverse source, $H_2(2\vec{k})$~ &   4.39    &  5.7     &  8.86      &  10.35      &  \numprint{122340}  \\
	  MLFSDA-$\omega$-$k,H_2(2\vec{k})$&   4.7    &  5.23     &   8.97     &  10.36     &   \numprint{3657}\\
	  $\abs{F_2(\vec{R},2k)}=R^3_z/R^4$&   4.7   &   5.23      &   8.97     &   10.36   &        506.83  \\
	  naive BPA                       &  4.87     &  5.51      &  8.98      &    10.36   &   225.82      \\
	\bottomrule
\end{tabular*}
\label{rus:pt_scat_res}
\end{table}

	\section{Conclusion}\label{concl}
	Starting from the forward scattering formulation of linearized inverse scattering problems with observation data available in a plane, several solution algorithms for the generation of spatial images of the underlying scattering distribution were derived, investigated, and compared. 
	The major focus was on an iterative solution method, which performs an inversion of the forward operator based on a discretization of the scattering distribution in the spatial frequency domain, but a corresponding direct inversion algorithm based on the spectral expansion of the forward scattering problem was also considered as an example of an improved adjoint imaging method. As a major contribution, the direct spectral inversion method with consideration of spectral low-pass filtering functions was also utilized to derive improved focusing operators for the standard spatial back-projection method.
	Both spectral algorithms work with fast operator evaluations based on the concepts of multi-level plane-wave representations as recently introduced in form of the multi-level fast spectral domain algorithm and as found in a similar manner in the multi-level fast multipole methods. In a series of test problems based on simulated and measured observation data, it was clearly found that the full operator inversion based inverse source solver produces considerably better imaging results than the adjoint imaging methods especially when the available observation data is not ideal, e.g., in the form of reduced sets of observation data or irregular sample locations. In all cases, imaging artifacts could be reduced by the introduction of spectral low-pass filtering functions.

	\section*{Acknowledgment}
	Funded by the European Union. Views and opinions expressed are however those of the author(s) only and do not necessarily reflect those of the European Union or European Innovation Council and SMEs Executive Agency (EISMEA). Neither the European Union nor the granting authority can be held responsible for them. Grant Agreement No:~101099491.

	\appendices

	\bibliographystyle{IEEEtran}
	\bibliography{Literature6}

	\begin{IEEEbiography}[{\includegraphics[width=1in,height=1.25in,clip,keepaspectratio]{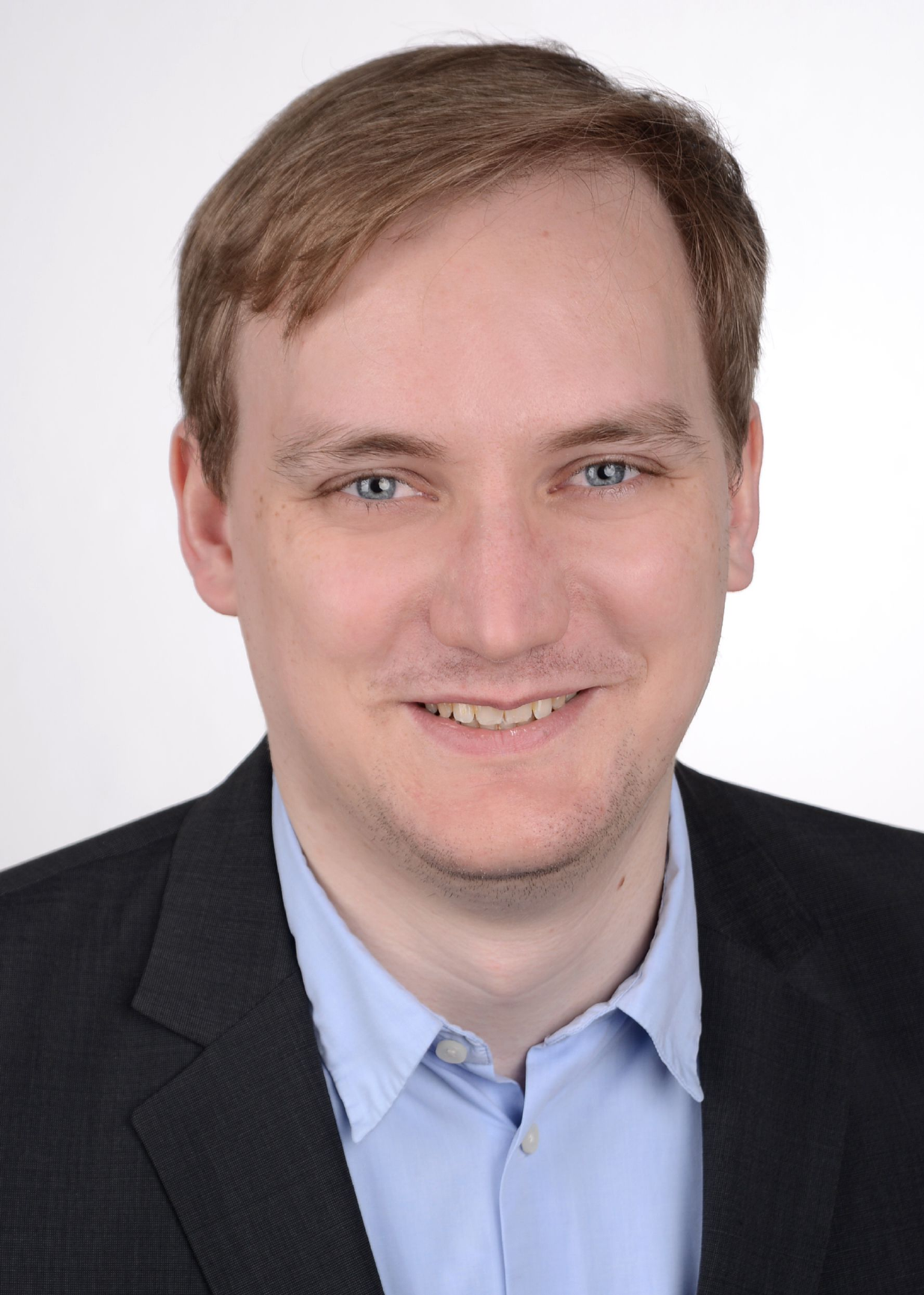}}]{Matthias Saurer}
    (Graduate Student Member, IEEE) received the M.Sc. degree in electrical engineering
	and information technology from the Technical University of Munich (TUM), Munich, Germany, in
	2021. Since 2021, he has been a Research Assistant with the Chair of High-Frequency Engineering,
	School of Computation, Information and Technology, TUM. His current research interests include
	the solution of inverse source problems, hierarchical
	multi-level acceleration, near-field scattering, and
	imaging problems. Moreover, in May 2023 he received the shared third price of the Young Scientist Award by the International
	Union of Radio Science (EMTS 2023).
	\end{IEEEbiography}
   
   \vspace{-4pt}

   \begin{IEEEbiography}[{\includegraphics[width=1in,height=1.25in,trim={0cm
   			6cm 4cm 0cm},clip,keepaspectratio]{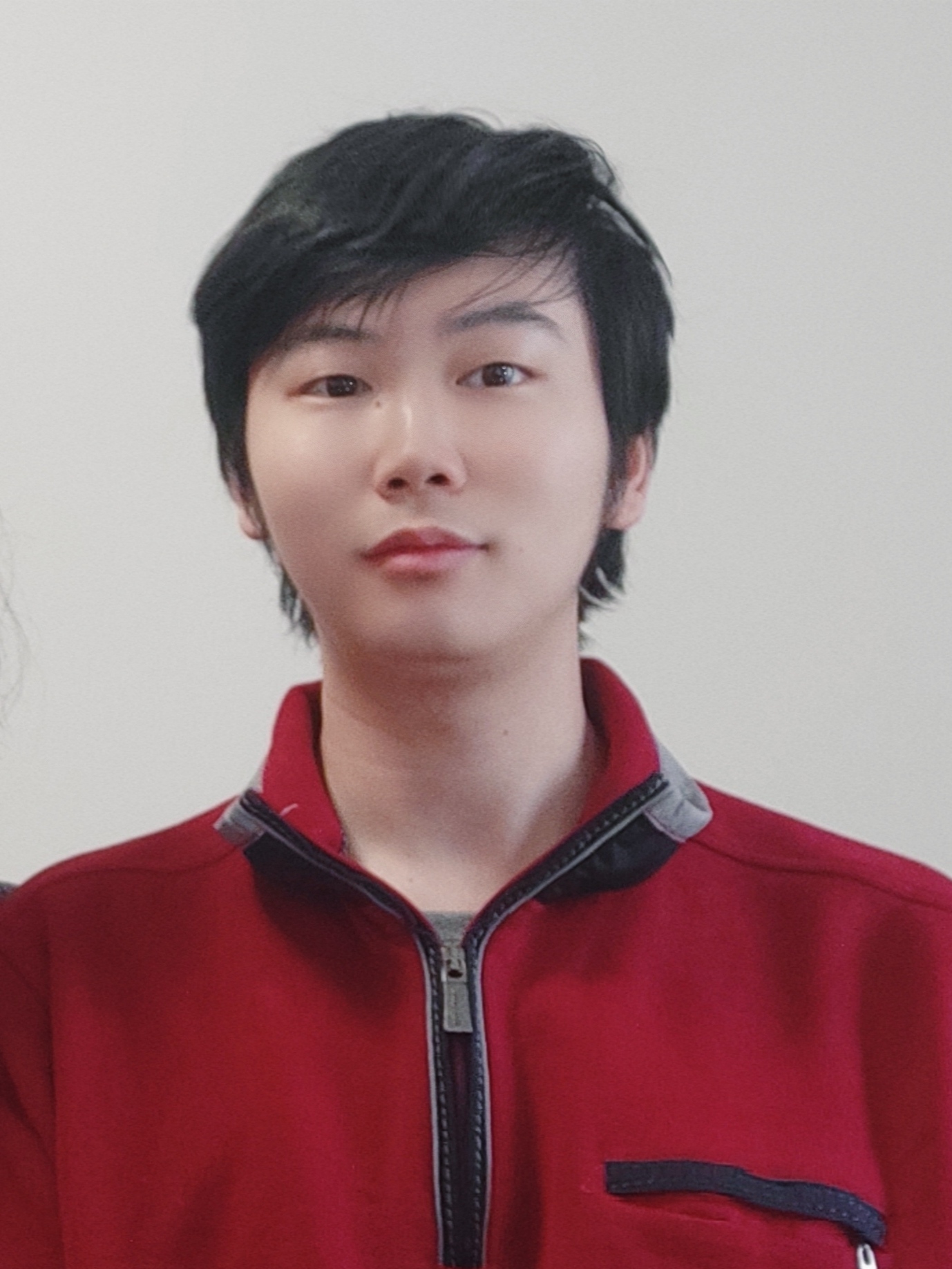}}]{Han Na} (Graduate Student Member,
   	IEEE) received the B.E. degree in electronic information engineering from
   	Beihang University, Beijing, China, in 2017, and the M.Sc. degree in
   	electrical engineering and information technology from the Technical
   	University of Munich (TUM), Munich, Germany, in 2020. Since 2021, he has been a Research Assistant with the Chair of
   	High-Frequency Engineering, School of Computation, Information and Technology, TUM. His research focuses on
   	electromagnetic ray tracing methods and microwave imaging. He has
   	co-authored several publications, including one in the \textsc{IEEE
   		Transactions on Antennas and Propagation}.	Mr. Na received the Best Paper Award at the 2023 Photonics \&
   	Electromagnetics Research Symposium (PIERS) and is a Student Member of the
   	IEEE Antennas and Propagation Society (AP-S).
   \end{IEEEbiography}
   
   \vspace{-4pt}
    
   \begin{IEEEbiography}
   	[{\includegraphics[width=1in,height=1.25in,clip,keepaspectratio]{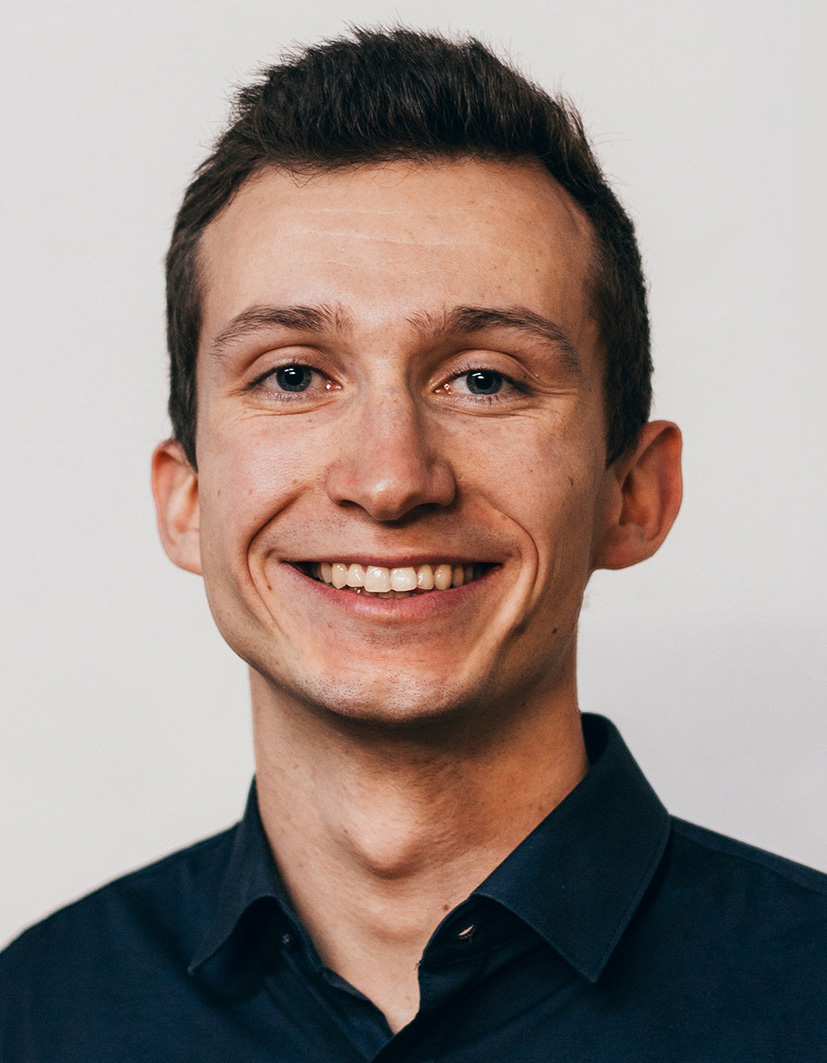}}]{Marius Brinkmann} (Graduate Student Member, IEEE) received the M.Sc. degree in electrical engineering from Florida Polytechnic University, Lakeland, FL, USA, in 2019, and the M.Sc. degree in electrical engineering and information technology from the Technical University of Munich (TUM), Munich, Germany, in 2021. Since 2022, he has been a doctoral candidate in a joint collaboration between the Chair of High-Frequency Engineering, School of Computation, Information and Technology, TUM  and Rohde \& Schwarz GmbH \& Co. KG, Munich, Germany. His current research interests include signal processing for high-resolution microwave imaging, near-field imaging, and calibration techniques. He is a recipient of the EuRAD 2022 Best Paper Award and the EuRAD 2024 Young Engineer Prize.
   \end{IEEEbiography}

   \vspace{-4pt}

   \begin{IEEEbiography}
	[{\includegraphics[width=1in,height=1.25in,clip,keepaspectratio]{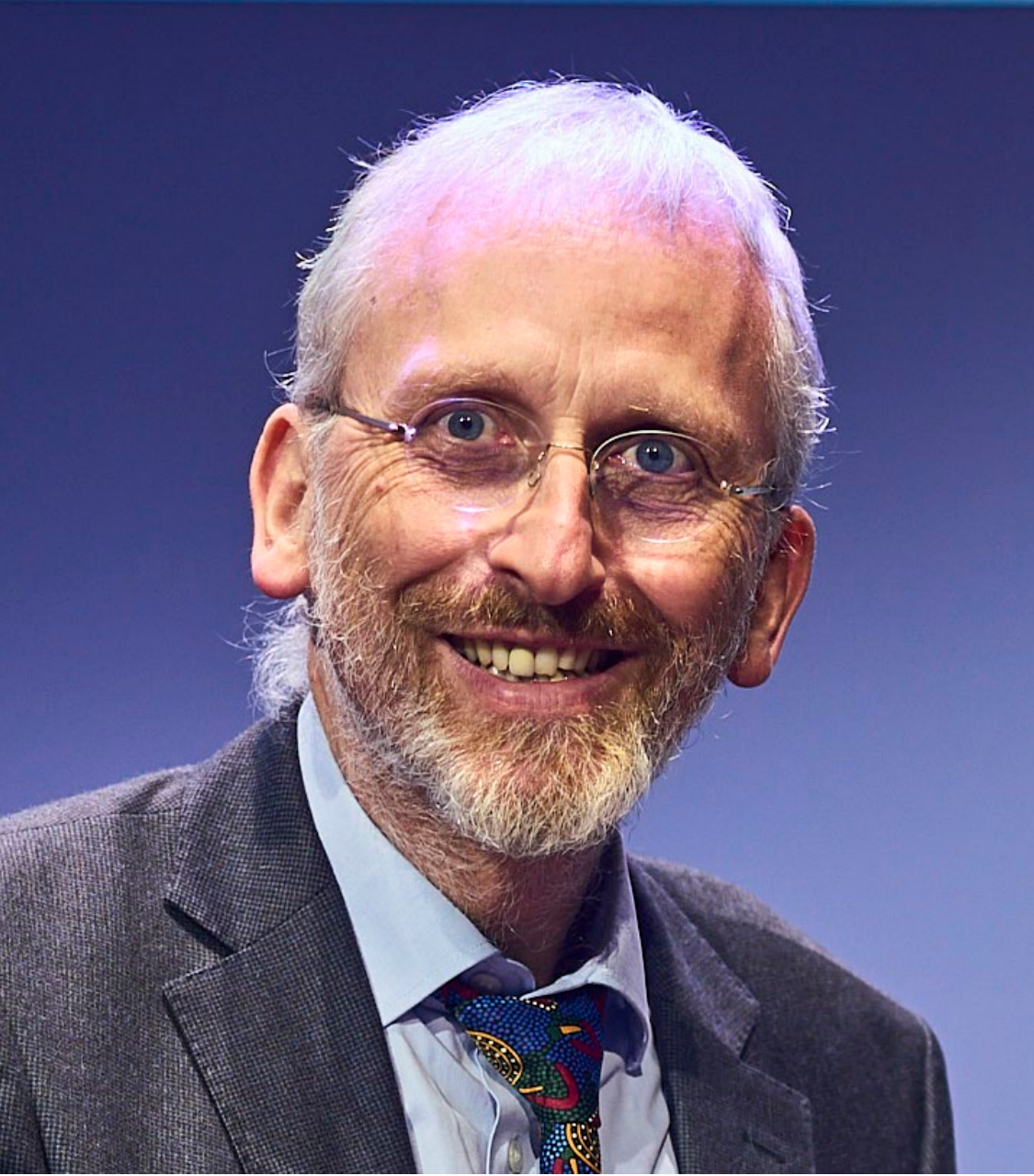}}]{Thomas F. Eibert} (Senior Member, IEEE)
	received the Dipl.-Ing.\,(FH) degree from Fachhochschule N\"urnberg, N\"urnberg, Germany, in 1989, the Dipl.-Ing.\;degree from Ruhr-Universit\"at Bochum, Bochum, Germany, in 1992, and the Dr.-Ing.\;degree from Bergische Universit\"at Wuppertal, Wuppertal, Germany, in 1997, all in electrical engineering.
	He is a Full Professor of high-frequency engineering with the Technical University of Munich, Munich, Germany, where he is currently also the Academic Program Director responsible for the study courses in the Professional Profile Electrical and Computer Engineering.
\end{IEEEbiography}

\end{document}